\documentclass[graybox]{svmult}

\usepackage{mathptmx}  
\usepackage{helvet}   
\usepackage{courier}  
\usepackage{type1cm}  
\usepackage{mathtools} 		
\usepackage{amssymb,amsmath,bbold}
\usepackage{makeidx}   
\usepackage{graphicx}  
\usepackage{float}			 
\usepackage{multicol}  
\usepackage[bottom]{footmisc}

\usepackage{amssymb,amsmath, bbold}
\usepackage{mathtools} 

\makeindex   

\newcommand{\ie}{{\it i.e.\/}}

\newcommand{\eg}{{\it e.g.\/}}

\newcommand{\nc}{\newcommand}  
\nc{\req}[1]{Eq.\,\ref{#1}}  
\nc{\rf}[1]{Fig.~\ref{#1}} 
\nc{\dsfrac}[2]{\displaystyle\frac{#1}{#2}}

\newcommand\blfootnote[1]{%
  \begingroup
  \renewcommand\thefootnote{}\footnote{#1}%
  \addtocounter{footnote}{-1}%
  \endgroup
}

\begin{document}

\title*{Probing QED Vacuum with Heavy Ions}
\author{Johann Rafelski, Johannes Kirsch, Berndt M\"uller,  Joachim Reinhardt,\\ \and Walter Greiner}
\institute{Johann Rafelski \at Department of Physics, The University of Arizona, Tucson, AZ 85721,\\ \email{rafelski@physics.arizona.edu}
\and Johannes Kirsch \at Frankfurt Institute for Advanced Studies, \email{jkirsch@fias.uni-frankfurt.de}
\and Berndt M\"uller \at Department of Physics, Duke University, Durham, NC 27708-0305
\and Joachim Reinhardt \at Institut f\"ur Theoretische Physik, Goethe-Universit\"at Frankfurt
\and Walter Greiner \at Frankfurt Institute for Advanced Studies}
%
%
\authorrunning{J.Rafelski,\,J.Kirsch,\,B.M\"uller,\,J.Reinhardt,\,and\,W.Greiner}
\maketitle
\vskip -1.5cm 
\abstract{We recall how nearly half a century ago the proposal was made to explore the structure of the quantum vacuum using slow heavy-ion collisions. Pursuing this topic we review the foundational concept of spontaneous vacuum decay accompanied by observable positron emission in heavy-ion collisions and describe the related theoretical developments in strong fields QED.
}\blfootnote{\it \hspace*{-0.1cm}Presented by JR at the International Symposium on \lq\lq New Horizons in Fundamental Physics: From Neutrons Nuclei via Superheavy Elements and Supercritical Fields to Neutron Stars and Cosmic Rays,\rq\rq\ held to  honor Walter Greiner on his 80th birthday at Makutsi Safari Farm, South Africa, November 23-29, 2015.}

\section{The Beginning}\label{Sec1}
The physics field of QED in strong fields and vacuum structure was born in 1929 when Oscar Klein~\cite{[11]} discovered what we call today the \lq Klein Paradox\rq. For the following four decades this field remained an academic curiosity. It surfaced as a research domain of acute interest about 50 years ago in the wake of the effort to create superheavy elements~\cite{WG1969}, as in pursuing this goal it became necessary to explore the physical properties of atomic nuclei of charge $Z >137$. This meant that we had to understand the physics of strongly bound relativistic electron eigenstates. 

The research program we address here began in the fall of 1968 when the precise quantitative solutions of the Dirac equation for finite superheavy nuclei were worked out~\cite{[7],[8]}, see Section~\ref{Sec21}. They showed the discrete eigenstate spectrum well known to be within the gap $-mc^2\le E_n\le mc^2$. These results offered a modern and quantitative view of earlier efforts; for a historical recount see Ref.\,\cite{Pop01}. 

In parallel to this awakening of interest in strong fields a new cycle of Frankfurt courses in theoretical physics had begun, with the first semester level theoretical physics lectures held by Walter Greiner. This new teaching program became the model for many other institutions. In the classroom were several students who soon shaped the tale of strong fields, among them two of the authors (BM and JR) who were soon attracted to work in strong fields physics.

By early 1970 the Strong Fields Frankfurt group was invited by Walter Greiner to a Saturday morning palaver in his office. In the following few years this was the venue where the new ideas that addressed the strong fields physics were born. At first the predominant topic was the search for a mechanism to stabilize the solutions of the Dirac equation, avoiding the ``diving'' of bound states into the Dirac sea predicted by earlier calculations~\cite{[7]}. However, a forced stability contradicted precision atomic spectroscopy data~\cite{JRDiplom71,[9],Rafelski:1973fm}. In consequence the group discussions turned to exploring the opposite, the critical field instability and the idea of spontaneous positron emission emerged. 

To best of our knowledge the first graphic rendering of the spontaneous positron production and related physical processes expected in supercritical field is the hand drawing \rf{Fig_m01} (p.\,79 Ref.\,\cite{JRDiplom71}). We see a bound-state deep within the negative energy continuum being filled by an electron $e^-$ barrier jump (tunneling), with the positron $e^+$ left outside of the potential well. Other processes that can occur are also shown: electron radiative capture into the supercritical state, and pair annihilation.
\begin{figure}[t]
\centering
\resizebox{0.65\textwidth}{!}{%
\includegraphics{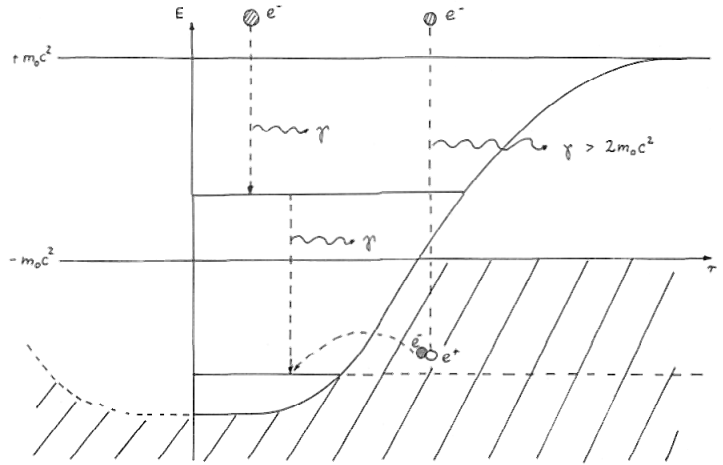}
}
\caption[]{Illustration showing the processes expected to occur for super bound unoccupied electron states including a virtual $e^-e^+$ pair separating with $e^-$ entering into the localized bound-state. Drawing of May 1971 by Helga Betz-Rafelski, from Ref.\,\cite{JRDiplom71}.~\label{Fig_m01}}
\end{figure}

The energy that the emitted positron would have is shown as connected to the location of this supercritical, deeply quasi-bound electron state. This novel process of positron production became soon known as the auto-ionization of positrons, an extension of the induced pair production process discovered by (Oscar) Klein~\cite{[11]}. The close connection to the spontaneous positron production was presented in the opening of the review of QED of strong fields by Rafelski, Fulcher and (Abraham) Klein~\cite{[3]}. We will discuss some of these aspects in Section~\ref{Sec22}.

The mathematical description to accompany these insights about vacuum instability and positron production was formalized a few months later~\cite{[22],[23]}, using the Fano embedding method~\cite{[40]}, see Section~\ref{Sec231}. We were able to show the presence of a \lq dived\rq\ resonance associated with a quasi-stable eigenstate, and soon after to redo this by solving exactly the Dirac equation for the scattering phases~\cite{[44]}, see Section~\ref{Sec232}. In the former Soviet Union Zel\rq dovich, Gershtein, and Popov~\cite{Pop01,Ger70,[13]} independently achieved a similarly complete \lq in principle\rq\ understanding of the physics of strong and supercritical fields. 

In historical perspective it is remarkable how quickly the key insights were gained both in Frankfurt and in Moscow. Knowing the dynamics of the Frankfurt effort from the inside, we can say that the strong fields group of Walter Greiner achieved within just a few months a full quantitative model allowing the localization of the quasi-bound-state, and the study in a quantitative manner of the resonance in the lower continuum as noted, see Section~\ref{Sec23}. 

Among the important developments was the understanding of the reformulation of quantum electrodynamics (QED) to accommodate the formation of the nonperturbative charged vacuum~\cite{[12]} state, described in Section~\ref{Sec3}. We will show that a supercritical domain in space spontaneously develops a localized charge cloud, and we will show how the back-reaction process stabilizes the charged vacuum state, Section~\ref{Sec333}.

The Frankfurt strong fields group worked out experimental observables, which required the study of the behavior of the Dirac electron-states present in quasi-molecular systems, see Section~\ref{Sec4}. When high $Z$-atoms collide such that the nuclear charge of both nuclei is supercritical but the nuclei only graze each other, the relativistic states can envelope both moving nuclei. Supercritical phenomena arise as transient effects in the collision process. The difficulty of this situation is that aside of the spontaneous vacuum decay we also encounter processes related to the time dynamics. A short summary of the 20 year long and inconclusive experimental effort is also presented in Section~\ref{Sec43}.

The highlight of this introduction to \lq\lq Probing QED Vacuum with Heavy Ions\rq\rq: the crucial technical step in the development of QED of strong fields has been the recognition that the spectrum of the Dirac equation in presence of supercritical fields contains a resonance in the negative energy continuum, a resonance continuously connected with the bound particle solutions reducing the strength of the potential from over-critical to sub-critical. For a strong field with somewhat less than the critical field strength, one can regard a vacancy in the $1s$-state as a bound positron state. Taking this view when the potential $V $ becomes supercritical the appearance of a positron at infinity is viewed simply as the delocalization of a bound positron state. Thus as $V$ is increased above $V_\mathrm{cr}$ spontaneous positron production will occur when and if the $1s$-state is empty. 

For full account of QED of strong fields written in the pioneering period 1970-80 see the three review articles Refs.\,\cite{[3],[1],[2]}, and our book~\cite{GMR85}. The account presented here focuses on the work that preceded the experimental effort. Some of material in this report is sourced from an unpublished review {\it Quantum Electrodynamics in Strong External Fields}~\cite{Kirsch:1981sf} whose theoretical sections remain timely and valid to the present day.

\section{Dirac equation and strong fields}\label{Sec2}

For an uninitiated reader of this report the first necessary insight is understanding why  we call the Coulomb potential that is capable of binding an electron by more than $2m_ec^2$ \emph{supercritical}. To answer this question let us  consider the electron-positron $e^-e^+$-pair production process. The minimum energy required is $2m_ec^2$. However, in the presence of a nucleus of charge $Ze$ it is possible that we do not require this vacuum  energy, since there is an electronic bound to the nucleus, and the binding reduces the pair energy threshold. 

The threshold for pair conversion of a $\gamma$-ray to an $e^-e^+$-pair in the presence of a nucleus is 
\begin{equation}
E_T^\gamma = m_ec^2+\epsilon_n\;,
\end{equation} 
where $\epsilon_n$ is the energy of the bound electron (always including its rest mass) in the eigenstate $n$. Considering the Pauli principle we recognize that this is only possible if such a state has not been occupied by another electron. The above energy balance for the $\gamma$-conversion to $e^-e^+$ pair implies the following statement:
\begin{quote} {\em When $\epsilon_n\to -m_ec^2$, the minimum energy required to create an $e^-e^+$-pair approaches zero: $E_T^\gamma\to 0$. At the critical point $\epsilon_n= -m_ec^2$ the energy of the ionized atom is equal to the energy of the atom with a filled $1s$-electron state and a free positron of nearly zero kinetic energy.}
\end{quote} 

It is important to consider carefully what happens if and when a metastable bound state $\epsilon_n\to \epsilon_R<-m_ec^2$ could exist. In such a situation the energy of a fully ionized atom without the $1s$-electron(s) is higher than the energy of an atom with \lq filled\rq\ K-shell and free positron(s). Thus a bare supercritical atomic nucleus cannot be a stable ground-state and therefore the neutral (speaking of electro-positron) vacuum cannot be a stable ground state as well.

We conclude that for super-critical binding where a quasi-state dives into the negative energy sea the supercritical bare atomic nucleus will spontaneously emit a positron $e^+$ (or two $e^+$ allowing for spin), keeping in its vicinity the accompanying negative charge which thus can be called the real vacuum polarization charge. The state that has an undressed atomic nucleus is the \lq neutral vacuum\rq\ (vacuum for electrons, positrons), and beyond the critical point is not the state of lowest energy. The new state of lower energy, called the charged vacuum~\cite{[12]}, is the dressed atomic nucleus; that is a nucleus surrounded by the real vacuum polarization charge which will be the conclusion of Section~\ref{Sec333}.

These introductory remarks show that the behavior of the spectrum of the Dirac equation is controlling the physics we are interested in. Thus in the remainder of this section following the footsteps of the early Frankfurt research results we address this important technical detail.

\subsection{Discrete spectrum in strong fields}\label{Sec21}
To describe electrons in an external electromagnetic field we use the Dirac equation for spin 1/2 particles, where we can adopt Coulomb gauge and thus for a (quasi)static electric field with $A^0 = V(r),\  \vec A=0$ 
\begin{equation}
H_\mathrm{D}\Psi_n(\vec{r})\ \equiv\ \left[\vec{\alpha}\cdot \vec{p}+\beta m+V(\vec{r})\right]\Psi_n(\vec{r})=E_n\Psi_n(\vec{r})\;.
\label{eq:eqO-2-2}
\end{equation}
In the Dirac representation the matrices $\vec{\alpha}$ and $\beta$ are given by
\begin{equation}
\vec{\alpha}=\left(\begin{matrix}
0 & \vec{\sigma} \\ \vec{\sigma} & 0
\end{matrix}\right)\ ,\ \ \ 
\beta=\left(\begin{matrix}
\mathbb{1} & 0 \\ 0 & -\mathbb{1}
\end{matrix}\right)\;,
\label{eq:eqO-2-3}
\end{equation}

Motivated by the later study of the potential generated by colliding heavy-ions we consider axially symmetric potentials 
\begin{equation}
V(\vec{r})=\sum_{l=0}^{\infty}V_l(r)P_l(\cos\theta)\,.\label{eq:eqO-2-6}
\end{equation}
For this case the most flexible approach~\cite{[39]} is based on a multipole expansion of the wave function
\begin{equation}
\Psi_\mu(\vec{r})=\sum_{\kappa}^{}\Psi_{\mu\kappa}(\vec{r})=\sum_{\kappa}^{}\left(\begin{matrix}
g_\kappa(r)\chi_\kappa^\mu \\ if_\kappa(r)\chi_{-\kappa}^\mu\end{matrix}\right)\;.\label{eq:eqO-2-4}
\end{equation}
$g_\kappa$(r) and $f_\kappa$(r) are the radial parts of the \lq large\rq\ and \lq small\rq\ components, respectively. The spinor spherical harmonics arise from coupling of spin-$1/2$ spinors $\chi_{\pm 1/2}$ with orbital eigenstates $ Y_{l}^{m}(\theta,\varphi)$
\begin{equation}
\chi^\mu_\kappa =\!\!\! \sum_{m=\pm\frac{1}{2}}C(l\ \frac{1}{2}\ j;\ \mu-m,\ m)\ Y_{l}^{\mu-m}\chi_m\,;
\qquad
 \kappa=
\begin{cases}
l\ &\mbox{for}\ \ j=l-1/2\;, \\ -l-1\ \ &\mbox{for}\ \ j=l+1/2\;.\end{cases} \label{eq:eqO-2-5}
\end{equation}
Here we introduced the eigenvalue $\kappa$ of the spin-orbit operator $\hat \kappa \equiv \beta (\vec{\sigma}\cdot \vec{l}+1)$. The total angular momentum $j = |\kappa|- 1/2$. The quantum number $\mu$ is the projection of $j$ on the symmetry axis which coincides for a two center potential of two separated nuclei with the connection line of the both field generating sources. 

The coupled radial equations read
\begin{subequations}
\begin{align}
\left(\frac{d}{dr} +\frac{\kappa +1}{r}\right) g_\kappa(r) &= -(E+m)f_\kappa(r) 
+\sum_{l,\kappa^\prime}^{}f_{\kappa^\prime}(r)V_l(r)\langle \chi^\mu_{-\kappa}|P_l|\chi^\mu_{-\kappa^\prime}\rangle \label{eq:eqO-2-7a}\\
\left(\frac{d}{dr}-\frac{\kappa -1}{r}\right)f_\kappa(r) &= -(E-m)g_\kappa(r) 
+\sum_{l,\kappa^\prime}^{}g_{\kappa^\prime}(r)V_l(r)\langle \chi^\mu_{\kappa}|P_l|\chi^\mu_{\kappa^\prime}\rangle \label{eq:eqO-2-7b}
\end{align}
\end{subequations}
For a spherically symmetric potential, i.e. $V(\vec r)=V_0(r)$, and $V_l = 0,\ \ \ l \geq 1$, the above simplifies to the usual result using $\langle \chi^\mu_{\pm\kappa}|P_0|\chi^\mu_{\pm\kappa^\prime}\rangle=\delta_{\pm\kappa,\pm\kappa^\prime}$. 
For the point nucleus $V_0(r) = -Z\alpha/r$, one finds the Sommerfeld relativistic fine structure formula:
\begin{equation}
E_{n\,\kappa} = m \left[1+\left(\frac{Z\alpha}{n-\kappa+\sqrt{\kappa ^2-(Z\alpha)^2}}\right)^2\right]^{-1/2}\!\!,\label{eq:eqO-2-10}
\qquad
n = 1,2,... \;.
\end{equation}
We note the singularity when $ Z\alpha \rightarrow |\kappa|$. For $ Z\alpha > |\kappa|$ some states have vanished, the remainder of the spectrum is incomplete and the Hermitian operator $H_\mathrm{D}$ ceases to be self-adjoint. This means that in a time evolution the probability of finding a particle is not conserved.

We sidestep the more mathematical discussion of possible self-adjoint extensions of $H_\mathrm{D}$ in presence of the singular $1/r$-potential. Instead, we explore physically motivated non-singular potentials that are obtained using a realistic nuclear charge distribution of a finite size nucleus of radius $R_N$
\begin{align}
V_0(r)=
\begin{cases}
-\frac{3}{2}\frac{Z\alpha}{R_N}\left(1-\frac{r^2}{3R^2_n}\right)\ &\mbox{for}\ \ 0\le r\le R_N\\[0.3cm]
-\frac{Z\alpha}{r} \ \ &\mbox{for}\ \ R_N < r< \infty\;.
\end{cases}\label{eq:eqO-2-11}
\end{align}
In order to include also the effect of the electron-electron interaction terms, Hartree-Fock-Slater calculations have been performed and the effects of vacuum polarization, see Section~\ref{Sec33}, and electron self energy, see Section~\ref{Sec334} were considered. Numerical results for the energy eigenvalues are shown in \rf{Fig_m02}. The eigen energy decreases monotonically as the nuclear charge $Z\alpha$ increases. 

None of the eigenvalues, or the wave functions in \rf{Fig_m02} exhibit any unusual behavior at $Z\alpha = 1$. The points at which the individual levels join the lower continuum are specific to each state. The critical $Z = Z_\mathrm{cr}$ value where the 1s level joins the lower continuum is $Z_{\mathrm{cr}}(1s_{1/2})\simeq 171.5$. The $2p_{1/2}$-state joins the lower continuum at $Z_{\mathrm{cr}}(2p_{1/2}) \simeq 185.5$. These values depend significantly on the assumed form of nuclear charge distribution, especially the nuclear radius $R_N(Z,A)$.  

\begin{figure}
\centering
\resizebox{0.7\textwidth}{!}{%
\includegraphics{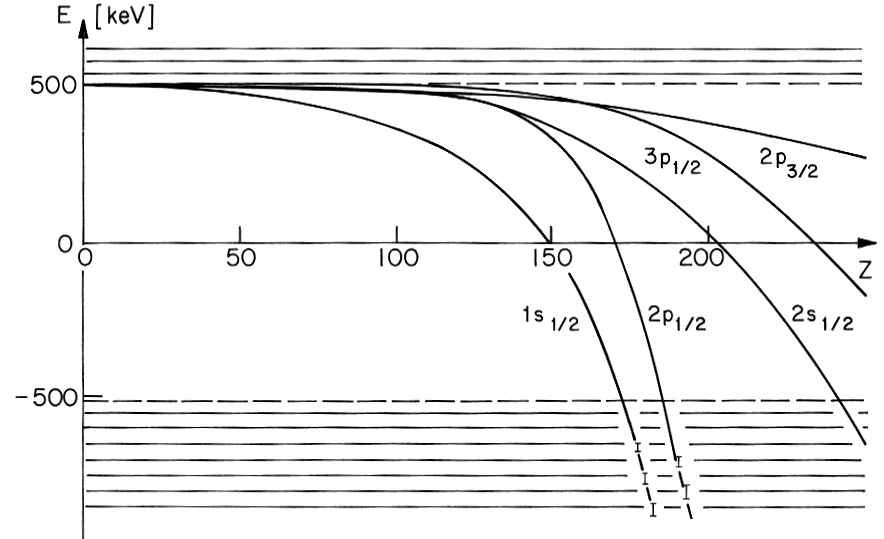}
}
\caption[]{The energies of the strongly bound Dirac atomic states versus the nuclear charge $Z$. 
}\label{Fig_m02}
\end{figure}

\subsection{Klein\rq s paradox}\label{Sec22}
In preparation to introduce the description of spontaneous $e^+$-production we turn next to the discussion of strong fields physics long ago begun with the paper by (Oscar) Klein~\cite{[11]}. We consider continuum states of the Dirac equation of an electron with momentum $p$ and energy $\epsilon=\sqrt{p^2+m^2}$, with spin up, that is incident from the left on an electrostatic square-well barrier $V_0> 0$. The discontinuous form of the potential requires that region I outside the potential well and region II inside the potential well be treated separately. Since $H_D$ is a first order differential form, at the barrier all the components of the Dirac spinor, but not their derivatives, need to be continuous.

In region I the plane wave solution of the free Dirac equation is
\begin{equation}
\Psi_I(z)=ae^{ipz} \left(\begin{matrix}
1 \\ 0 \\ p/(\epsilon +m) \\0 \end{matrix}\right) +be^{-ipz} \left(\begin{matrix}
1 \\ 0 \\ -p/(\epsilon +m) \\0 \end{matrix}\right)\;,\label{eq:eqO-10-1}
\end{equation}
where the second part of the wave function describes the reflected wave. The incident current $j_i$ is
\begin{equation} 
j_i=\Psi_i^+\alpha_3\Psi_i=2p\frac{|a|^2}{\epsilon+m}\;. 
\end{equation}
The form of the wave function in region II depends upon the magnitude of the potential strength. For values of potential that are small  $|V|<2m$, the situation is analogous to non-relativistic quantum mechanics; nothing can penetrate the barrier and one easily finds that the ratio of the reflected current to the incident current is $|b|^2/|a|^2=1$. 

Let us now consider what happens if $V_0$ is increased to values seen on left in \rf{Fig_m03}. The wave function must be written as:
\begin{equation}
\Psi_{II}(z)=de^{ip^{\prime \prime} z} \left(\begin{matrix}
1 \\ 0 \\ p^{\prime \prime} /(\epsilon +m -V_0) \\0 \end{matrix}\right)\ \ ,\ \ \epsilon +m < V_0 \;,\label{eq:eqO-10-4}
\end{equation}
where $p^{\prime \prime } =\sqrt{(V_0-\epsilon)^2-m^2}$. The continuity condition leads to
\begin{align}
d/a&=2/(1-\Gamma^\prime) \label{eq:eqO-10-5} \\
b/a&=(1+\Gamma^\prime)/(1-\Gamma^\prime) \nonumber
\end{align}
with
\begin{equation}
\Gamma^\prime =[(\epsilon +m)(V_0-\epsilon +m)]^{1/2}[(\epsilon -m)(V_0-\epsilon -m)]^{-1/2} \;.\nonumber
\end{equation}
The transmitted current is equal to $2p^\prime|d|^2/(\epsilon+m-V_0)$, which is
negative, and the magnitude of the reflected current is larger
than that of the incident current. The transmission coefficient,
which is the ratio of the transmitted current to the incident
current, is given by
\begin{equation}
T=-\frac{4\Gamma^\prime}{(1-\Gamma^\prime)^2} \;.\label{eq:eqO-10-6}
\end{equation}

\begin{figure}[t]
\centering
	\resizebox{0.95\textwidth}{!}{
	\includegraphics{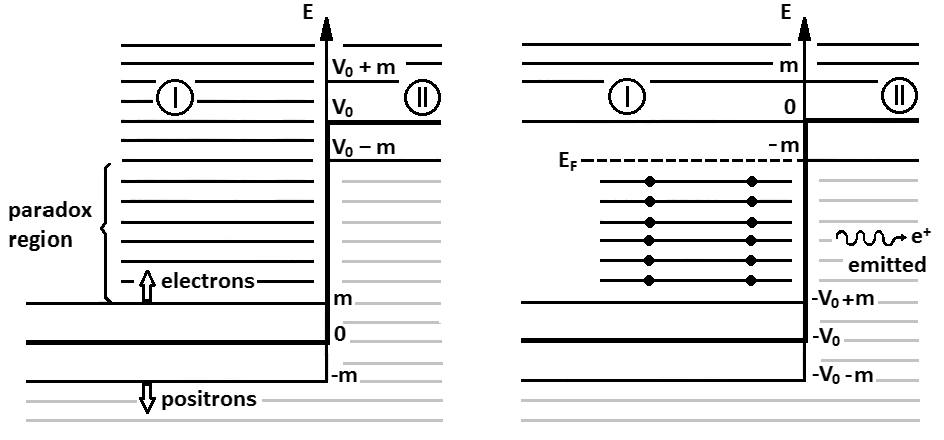}
	}
\caption[]{On left: For $+m< \epsilon < V_0-m\ (V_0> 0)$ the transmission coefficient of electrons impinging in region I on potential $V_0$ is
larger than unity. On right: Negative potential step as created by a large \lq nucleus\rq.
All states with $-V_0+m< \epsilon < -m$ in region I are filled
by spontaneous positron emission into region II.} \label{Fig_m03}
\end{figure}

We note that the transmitted current may be much larger than the incident current. This is the key finding of  Oscar Klein~\cite{[11]}. The domain of eigenstates for which this behavior occurs is shown in \rf{Fig_m03} on left. Only in the context of a single-particle interpretation does this result appear paradoxical. When one appreciates that electrons and positrons are inextricably connected in the Dirac theory, it is natural to identify the negative current in region II with the appearance of positrons. However in 1929, at the time Klein wrote his paper, positron interpretation of the second continuum had not as yet been recognized. The increase of the reflected current over the incident current is necessary to conserve charge. The reflected current plus the transmitted current is always equal to the incident current.

To make contact with situation of large nuclei  we redefine the reference point of the energy: we consider the potential step to be strongly attractive in region I $(0\rightarrow -V_0)$ and zero in region II $(V_0\rightarrow 0)$, see \rf{Fig_m03} on right. Then we expect that within the finite or infinite range of the attractive potential all supercritical states are spontaneously filled with \lq electrons\rq while the positrons are emitted to infinity. Now Klein\rq s gedankenexperiment consists of scattering positrons off the (filled) attractive potential well. Since no final states are available for electrons within the well, we find that the transmission coefficient vanishes, no other particles can be transmitted into the region of the potential. We note that the potential step must be determined in a self-consistent manner as we describe in Section~\ref{Sec333}: the background charge and the charge of the filled states must be combined to produce the potential barrier under consideration.

\subsection{Quasi-bound state in negative energy continuum}\label{Sec23}
\subsubsection{Fano resonance}\label{Sec231}
We now describe the quasi-bound states  with $E<m$. A semi-analytical solution is obtained once we approximate the potential 
\begin{equation}
V(r,Z) =Z\bar U(r;Z)\simeq Z\ U(r)\,. \label{eq:eqO-2-13}
\end{equation}
Within the range of atomic nuclei considered by us, $ 170< Z< 200$, the quantity $\bar U(r;Z)$ is weakly dependent on $Z$ via the radius of the nuclear charge distribution. We now use the fact that we know the solution to our problem for $Z = Z_\mathrm{cr}$ and diagonalize $H_\mathrm{D}(Z=Z_\mathrm{cr}+Z^\prime )$ in the basis of eigenstates given by $H_\mathrm{D}(Z_\mathrm{cr})$. Let $\Phi$ be the $1s$-state eigenfunction for $Z = Z_\mathrm{cr}$, i.e.
\begin{equation}
H_\mathrm{D}(Z_\mathrm{cr})\Phi=E_0\Phi\ \simeq\ -m\Phi \label{eq:eqO-2-14}
\end{equation}
and $\Psi_E$ be the orthogonal $s$-continuum wave functions with $E<-m$
\begin{equation} 
H_\mathrm{D}(Z_\mathrm{cr})\Psi_E = E\Psi_E\;,\qquad \label{eq:eqO-2-15} 
\langle \Psi_{E^\prime}|\Psi_{E^{\prime\prime}}\rangle=\delta(E^\prime-E^{\prime\prime})\,.
\end{equation} 
$\Phi$ and all $\Psi_E$ serve as a  basis for our diagonalization procedure. We neglect the small contribution from the higher bound $ns$ states $ n>1 $ which are separated by more than 500 keV from the $1s$-bound-state. We will need the matrix elements of $H_\mathrm{D}(Z_\mathrm{cr}+Z^\prime )$ in our truncated basis
\begin{subequations}
\begin{align}
\langle \Phi|H_\mathrm{D}(Z_\mathrm{cr}+Z^\prime )|\Phi\rangle &= E_0+\Delta E_0\;,\qquad \label{eq:eqO-2-16}
 \Delta E_0 =Z^\prime \langle \Phi|U(r)|\Phi\rangle\;,\\[0.2cm] 
\langle \Psi_E|H_\mathrm{D}(Z_\mathrm{cr}+ Z^\prime )|\Phi\rangle &= V_E\;,\qquad \qquad \qquad \label{eq:eqO-2-18}
 V_E = Z^\prime \langle \Psi_E|U(r)|\Phi\rangle\;,\\[0.2cm]
\langle \Psi_{E^{\prime\prime} }|H_\mathrm{D}(Z_\mathrm{cr}+ Z^\prime )|\Psi_{E^\prime}\rangle &= 
E^\prime\delta (E^{\prime\prime} -E^\prime) 
+ U_{E^{\prime\prime} E^\prime}\;, \ 
\label{eq:eqO-2-19}  
 U_{E^{\prime\prime} E^\prime} = Z^\prime\langle \Psi_{E^{\prime\prime} }| U(r)|\Psi_{E^\prime}\rangle\;.
\end{align}
\end{subequations}
The matrix elements $U_{E^{\prime\prime} E^\prime}$ describe the rearrangement of the continuum states under the additional potential $U(r)$. For small $Z^\prime$ this may be neglected since its influence upon the ls bound-state is a second order effect.
Our aim is to find $\tilde{\Psi}_E$, a continuum solution to the Dirac equation for $Z> Z_\mathrm{cr}$ in terms of the truncated basis, that is to solve 
\begin{equation}
H_\mathrm{D}(Z_\mathrm{cr}+Z^\prime )\ \tilde{\Psi}_E\ = \ E\ \tilde{\Psi}_E\;,
\label{eq:eqO-2-20}
\end{equation}
where the continuum functions $\tilde{\Psi}_E$ are normalized in the usual way
\begin{equation}
\langle \tilde{\Psi}_E^\prime \ |\ \tilde{\Psi}_E\rangle =\delta(E^\prime\ -\ E)\;.\nonumber
\end{equation}
Following Fano method we expand $\tilde{\Psi}_E$ within the space spanned by the truncated basis comprising one bound state and one continuum
\begin{equation}
\tilde{\Psi}_E=a(E)\ \Phi\ +\ \int_{|E^\prime|> m}b_{E^\prime}(E)\Psi_{E^\prime}dE^\prime\;.
\label{eq:eqO-2-21}
\end{equation}
The coefficients $a(E)$ and $b_{E^\prime}(E)$ are readily determined. We are mainly interested in the effects on the bound-state $\Phi$ and find~\cite{[22]}
\begin{equation}
|a(E)|^2=\frac{|V_E|^2}{[E-(E_0+\Delta E_0)-F(E)]^2+\pi^2|V_E|^4}\;,
\label{eq:eqO-2-22}
\end{equation}
where $F(E)$ is the principal value integral
\begin{equation}
F(E)=\mathbb{P}\ \int_{|E^\prime|> m}^{}dE^\prime\frac{|V_{E^\prime}|^2}{E-E^\prime}\,.
 \label{eq:eqO-2-23}
\end{equation}
The quantity $|a(E)|^2$ is the probability that an electron bound in $\Phi$ is embedded in $\tilde{\Psi}_E$ as the additional charge $Z^\prime$ is \lq\lq switched on\rq\rq. The quantity $|a(E)|^2$ has resonance Breit-Wigner shape 
\begin{equation}
|a(E)|^2=\frac{1}{2\pi}\frac{\Gamma}{[E-(E_0+\Delta E_0)]^2+\Gamma^2/4}\;,\qquad \label{eq:eqO-2-24}
\Gamma=2\pi\ |V_E|^2\ \simeq\ Const.
\end{equation}
with the resonance peaked around $E_0+\Delta E_0$. Writing \req{eq:eqO-2-24} we have neglected $F(E)$ with respect to $\Delta E_0$ and introduced $\Gamma$ which is possible when $V_E$ does not depend too strongly on the
energy $E$ -- this is the case once the state dived a bit \eg\ $Z^\prime  >  3$. Since we have chosen $E_0\simeq-m$ 
\begin{equation}
\Delta E_0=Z^\prime \langle \Phi|U(r)|\Phi\rangle \ \equiv\ -Z^\prime \delta\, \label{eq:eqO-2-25}
\end{equation}
describes the energy shift of the bound $1s$-state due to the
additional charge $Z^\prime$\;. The width $\Gamma$ of the resonance is
\begin{equation} 
\Gamma\ =\ 2\pi|V_E|^2=2\pi|Z^\prime \langle \Psi_E|U(r)|\Phi\rangle |^2\equiv\ Z^{\prime \,2}\gamma \;. \label{eq:eqO-2-26} 
\end{equation} 
Calculations show that~\cite{[23]}
\begin{equation}
\delta\ \simeq\ 30\;\mathrm{keV}\;,\qquad \gamma\ \simeq\ 0.05\;\mathrm{keV}\;.
\label{eq:eqO-2-27}
\end{equation}
We may explicitly show the $Z^\prime$-dependence of \req{eq:eqO-2-24}:
\begin{equation}
|a(E)|^2=\frac{1}{2\pi}\frac{Z^{\prime \,2}\gamma}{[E+m+Z^\prime \delta]^2+Z^{\prime\,4}\gamma^{\,2}/4}\;.
 \label{eq:eqO-2-28}
\end{equation}
From \req{eq:eqO-2-28} we learn that the bound-state $\Phi$ \lq\lq dives\rq\rq\ into the negative energy continuum for $Z> Z_\mathrm{cr}$ proportional to $Z^\prime = (Z-Z_\mathrm{cr})$. At the same time it obtains a width $\Gamma_E$ within the negative energy continuum proportional to $Z^{\prime \,2} = (Z-Z_\mathrm{cr})^2$. 

\subsubsection{Scattering phases}\label{Sec232}
These physics illuminating results can be also obtained by directly solving the Dirac equation for phase shifts of the lower continuum wave functions, a procedure which is also required for large $ Z^\prime$~\cite{[23],[41]}. From the ratio of the radial functions at the nuclear surface a phase shift $\delta$ is determined. The results for $\sin^2(\delta-\delta_0)$ are represented in \rf{Fig_m04}. The background phase $\delta_0$ was calculated using a weaker potential of a nucleus with three fewer protons, \ie\ for $Z_0=Z-3$.
\begin{figure}[t]
\centering
\resizebox{0.7\textwidth}{!}{%
\includegraphics[width=\columnwidth]{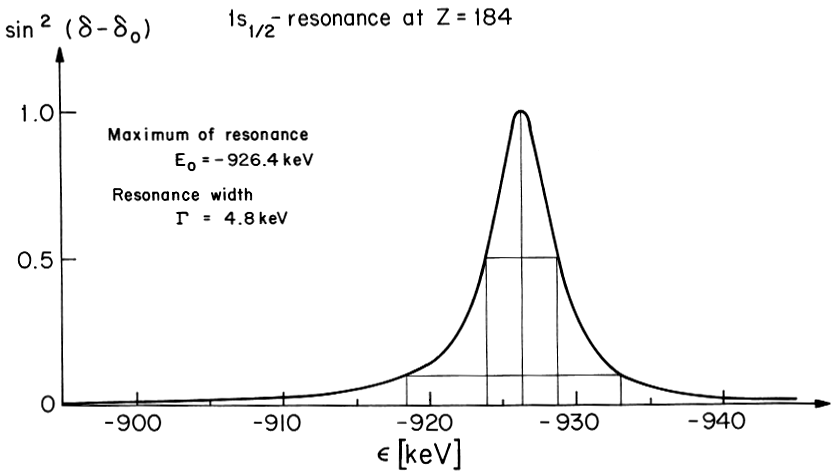} 
}
\caption[]{The energy dependence of $\sin^2(\delta-\delta_0)\equiv |a(E)|^2$ in an supercritical electrostatic potential $Z=184$.}\label{Fig_m04}
\end{figure}
The resonance in \rf{Fig_m04} is centered at $\epsilon=E_0=-926keV$ and the full width at half maximum is $\Gamma=4.8keV$ in excellent agreement with the Fano method results we presented just before. The results for the resonance energy $\epsilon$ of the $1s_{1/2}$ and $2p_{1/2}$ resonances as functions of the nuclear charge are shown in \rf{Fig_m05}. 
\begin{figure}[t]
\centering
\resizebox{0.56\textwidth}{!}{%
\includegraphics{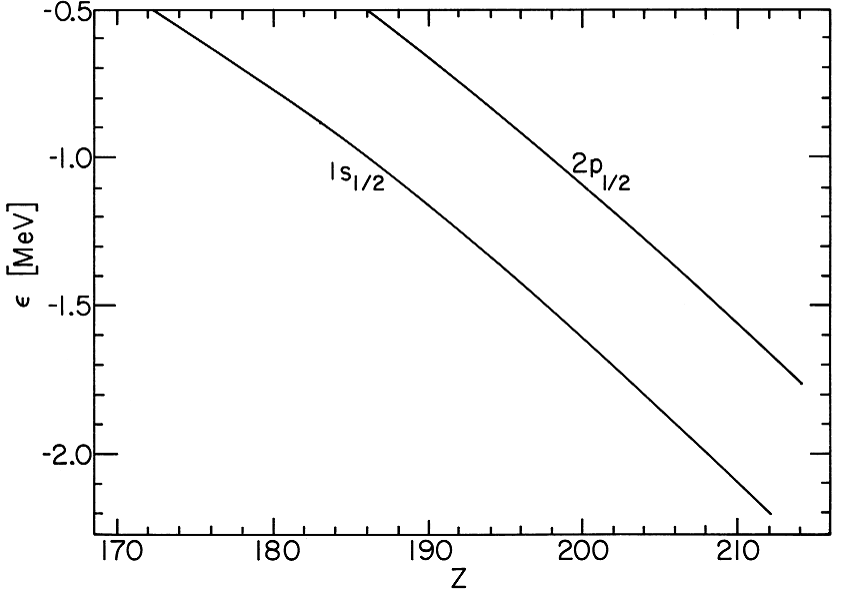}
}
\caption[]{The values $\epsilon$ of the $1s_{1/2}$ and $2p_{1/2}$ resonances as functions of the nuclear charge $Z>Z_\mathrm{cr}$.}\label{Fig_m05}
\end{figure}
The value $\epsilon$ and the width $\Gamma$ of the resonance are found to follow in this domain a simple parametric form
\begin{equation} 
\epsilon  \simeq -(Z-Z_\mathrm{cr})\delta-(Z-Z_\mathrm{cr})^2\tau\;, \qquad
\label{eq:eqO-2-32}
 \Gamma \simeq (Z-Z_\mathrm{cr})^2\gamma\;, 
\end{equation} 
The expression for the resonance location $\epsilon$ obtained for $Z>Z_\mathrm{cr}$ is also describing the location of the bound state for $Z< Z_\mathrm{cr}$. The expression for the width $\Gamma$ is applicable only if $Z> Z_\mathrm{cr}$. Moreover, for values of $Z$ nearer $Z_\mathrm{cr}$, it is necessary to include a dampening factor allowing that the probability of finding low energy positrons near the nucleus is small when $Z\sim Z_\mathrm{cr}$. Values for $\delta$, $\tau$ and $\gamma$ are listed in Table \ref{table21}. 
\begin{table}[H]
\centering
\caption{Parameters for the $1s_{1/2}$ and $2p_{1/2}$ resonances.}
\label{table21} 
\begin{tabular}{p{2cm}p{2cm}p{2cm}}
\hline\noalign{\smallskip}
& $1s_{1/2}$ & $2p_{1/2}$ \\
\noalign{\smallskip}\svhline\noalign{\smallskip}
$Z_{\rm cr}$ & 171.5 & 185.5 \\
$\delta$ (keV) & 29.0 & 37.8 \\
$\tau$ (keV) & 0.33 & 0.22 \\
$\gamma$ (keV) & 0.04 & 0.08 \\
\noalign{\smallskip}\hline\noalign{\smallskip}
\end{tabular}
\end{table}
The motivation for writing the results in the form of  \req{eq:eqO-2-32} is to make contact with the previous approach for the calculation of the resonance parameters and to make simple parametric equations available for calculations. 

Thus we have in full described how for $Z> Z_\mathrm{cr}$ a quasi-bound state can be found embedded amongst the continuum states. We have seen that as the proton number of a nucleus with $Z< Z_\mathrm{cr}$ is steadily increased, the energy of K-shell electrons  $E_{1s}$ is decreased until at $Z=Z_\mathrm{cr}$ it reaches $E_{1s}=-m_e$. During this process the spatial extension of the K-shell electron charge distribution is also decreasing, \ie\ the bound-state wave function becomes more and more localized. 

When $Z$ grows beyond $Z_\mathrm{cr}$ the bound $1s$-state ceases to exist. But that does not mean that the K-shell electron cloud becomes delocalized. Indeed, according to \req{eq:eqO-2-21} the bound state $\Phi$ is shared by the negative energy continuum states in a typical resonance manner over a certain range of energy seen in \req{eq:eqO-2-22}. The negative energy continuum wave functions become, due to the bound-state admixture, strongly distorted around the nucleus. 

This additional distortion of the negative energy continuum due to the bound-state can be called \emph{real charged vacuum polarization}~\cite{[12]}, because it is caused by a real electron state which joined the \lq\lq ordinary vacuum states\rq\rq , \ie\ the negative energy continuum. The charge densities induced by the continuum states superpose to form an electron cloud of K-shell shape. 
The total probability (up to spin degeneracy) for finding the $1s$-electron state $\Phi$ in any of the continuum states is $(\gamma\ll \delta)$:
\begin{equation}
\int_{-\infty}^{-m}\ dE\ |a(E)|^2=1\;.\label{eq:eqO-2-29}
\end{equation}
The K-shell electron cloud remains spatially localized in $r$-space. However, it obtains an energy width $\Gamma$. 

This can be illustrated in the following way: Consider the Dirac equation with the cut-off Coulomb potential inside a finite sphere of radius $a$. Certain boundary conditions on the sphere have to be fulfilled. In this way the continuum is discretized, see \rf{Fig_m06}.  On the left the situation at $Z=Z_\mathrm{cr}$, \ie\ before diving is shown. After diving we see on right that the $1s$-bound-state is spread over it. In that sense the K-shell still exist, but electrons are spread out energetically. An observable consequence would be that an induced $1s\to 2p$-excitation by $\gamma$-absorption would acquire an additional width, the spreading width.
\begin{figure}[t]
\centering
\resizebox{0.69\textwidth}{!}{%
\includegraphics{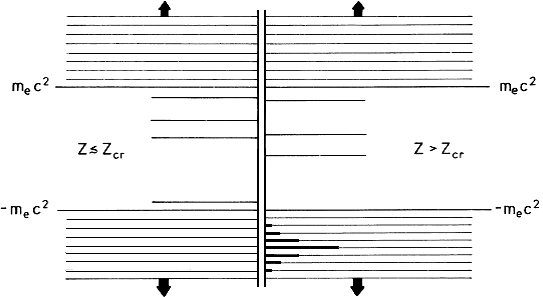}
}
\caption[]{On left spectrum before, and on right after diving showing the spreading of the formerly bound-state  over the negative energy continuum states.}\label{Fig_m06}
\end{figure}

The situation is different when the $1s$-bound-state is empty (ionized) during diving and $Z$ is increased beyond $Z_\mathrm{cr}$. Then, on grounds of charge conservation, one of the resulting continuum states $\Psi$ has to be empty, \ie\ a positron escapes. The observed kinetic energy spectrum of the escaping positron  has a Breit-Wigner type spectrum given by \req{eq:eqO-2-24}. Thus the width $\Gamma$ that describes the lifespan of quasi-bound state is also the positron escape width. 

The probability per unit time for emission of positrons in the energy interval $dE$ is given by Fermi\rq s \lq Golden Rule\rq:
\begin{equation}
p(E)\ dE=\frac{2\pi}{\hbar}\ |\langle \Phi |H_{int}|\Psi _E\rangle |^2\;\rho(E)\;dE\;.
\label{eq:eqO-2-30}
\end{equation}
The continuum states $\tilde{\Psi}_E$ are normalized to $\delta$-functions and the probability for finding the bound-state at the energy $ E$ is according to \req{eq:eqO-2-21} given by $\rho(E)dE = |a(E)|^2dE$, and $H_{int}=Z^\prime U(r)$. Hence the decay probability of the empty K-shell embedded in the negative continuum, \ie\ rate per unit time, is, using \req{eq:eqO-2-24}
\begin{equation}
p(E)dE=\frac{1}{2\pi}\ \frac{\Gamma_EdE}{[E-(E_0+\Delta E_0)]^2+\Gamma_E^2/4}\ \frac{\Gamma_E}{\hbar}\;.
\label{eq:eqO-2-31}
\end{equation}

This decay will be interpreted (see below) as the \emph{decay of the normal, neutral vacuum into a charged vacuum} (charge 2 for $Z > 172$) in supercritical fields. The normal vacuum state is absolutely stable up to $Z=Z_\mathrm{cr}$ and becomes unstable (spontaneous decay) in supercritical fields. Only the charged vacuum (after two positrons were emitted) is stable in supercritical fields. The vacuum proceeds to become higher charged as the supercritical fields (supercritical charge) are further increased. The above results can easily be generalized to several supercritical states embedded in the negative continuum. Ultimately, we will present a description that incorporates the screening of the source charge by the real charge density of the vacuum in Section \ref{Sec333} .

\section{Quantization of the Dirac field and the (charged) vacuum state}\label{Sec3}
\subsection{Canonical quantization}\label{Sec31}
Our study of single-particle eigenstates of the Dirac equation combined with physical intuition showed that in presence of strong fields there is a nonperturbative instability leading to positron production. Whenever particles are produced it is necessary to consider the 2nd quantization, which we now introduce using the canonical approach~\cite{[52]} and address the nature of the ground-state~\cite{[12],[64a],[64b]}. The abrupt change of the ground-state (phase transition) for sufficiently strong potentials, accompanied by pair-creation will be found as the main consequence. 

We introduce a Heisenberg operator $\hat{\Psi}(\vec{x},t)$ that acts in the Fock space of state vectors. The basic meaning of the operator $\hat{\Psi}(\vec{x},t)$ is that it annihilates a particle or creates an antiparticle at time $t$ at the space-point $\vec{x}$. In many cases it is more practical to characterize particles not by position $\vec{x}$ but by a normalizable stationary wave function $ {\psi}(\vec{x})$. Given a complete set of such functions spanning a Hilbert space, we can divide it into two subsets: one describing particles and one describing antiparticles where we shall denote the sets symbolically by \lq\lq $n> \mathrm{F}$\rq\rq\ and \lq\lq $n< \mathrm{F}$\rq\rq, respectively. Accordingly we write:
\begin{equation}
\hat{\Psi}(\vec{x},t=0)=\sum_{n> \mathrm{F}}^{}\hat{b}_n\psi_n(\vec{x}))\ +\ \sum_{n< \mathrm{F}}^{}\hat{d}^\dagger _n\psi_n(\vec{x})) \;.\label{eq:eqO-4-1}
\end{equation}
$\hat{b}$ annihilates a $e^-$ in the single-particle state $\psi_n$, $\hat{d}^\dagger$ creates a $e^+$  in state $\psi_n$. In this section we shall restrict ourselves to situations where the external potential is time-independent, allowed to assume that the functions $\psi_n$ are eigenfunctions of the single-particle Hamiltonian, \req{eq:eqO-2-2}. The conjugate operator to $\hat{\Psi}$ denoted by $\hat{\Psi}^\dagger$ is creating an electron and destructing a positron. An analogous decomposition is
\begin{equation}
\hat{\Psi}^\dagger (\vec{x},t=0)= \sum_{n> \mathrm{F}}^{}\hat{b}_n^\dagger \psi_n^\dagger (\vec{x})\ +\ \sum_{n< \mathrm{F}}^{}\hat{d}_n\psi_n^\dagger (\vec{x})\;. \label{eq:eqO-4-4}
\end{equation}

In terms of $\hat{\Psi}$ and $\hat{\Psi}^\dagger$ one can construct a Lagrangian operator implementing hermiticity and charge conjugation invariance by taking the properly symmetrized expression~\cite{Schwinger:1951nm}. One then finds the canonically conjugate momentum to $\hat{\Psi}$
\begin{equation}
\hat{\Pi}=
\frac{\partial {\cal L} (\hat{\Psi}, \dot{\hat{\Psi}})}{\partial \dot{\hat{\Psi}}}
=i\hat{\Psi}^\dagger\;.\notag
\end{equation}
The following equal-time anticommutation relations on the field operators are imposed:
\begin{align}\label{eq:eqO-4-5}
\{\hat\Psi(\vec x,0),\hat\Psi^\dagger(\vec x,0)\} &=\delta(\vec x-\vec x^\prime)
\;, \\
\{\hat\Psi(\vec x,0),\hat\Psi(\vec x,0)\}
&=\{\hat\Psi^\dagger(\vec x,0),\hat\Psi^\dagger(\vec x,0)\}=0 \;.
\nonumber 
\end{align}
The decompositions \req{eq:eqO-4-1} and \req{eq:eqO-4-4} lead to the relations
\begin{align}\label{eq:eqO-4-6}
\{\hat b_n,\hat b_m\}&=\{\hat b^\dagger_n,\hat b^\dagger_m\}
=\{\hat d_n,\hat d_m\}=\{\hat d^\dagger_n,\hat d^\dagger_m\}=0 \;,
\\ \nonumber
\{\hat b_n,\hat b^\dagger_m\}&=\{\hat d_n,\hat d^\dagger_m\}=\delta_{mn}
\;.
\end{align}
These equations must be completed by an equation that determines the dynamical time evolution of the field operators $\hat{\Psi}$, $\hat{\Psi}^\dagger$ that is $\hat b_n$, $\hat b_n^\dagger$, $\hat d_n$, $\hat d_n^\dagger$. It is convenient in the study of supercritical phenomena to work in the Heisenberg picture where the Fock-state-vector is time independent and the dynamics are determined by the operators according to Heisenberg\rq s equations of motion.

In order to be the generator of a unitary time-evolution the Hamiltonian must be constructed as a self-adjoint operator. This is achieved (for well-behaved potentials) by complete symmetrization with respect to the field operator, which renders $\hat{H}$ a Hermitian operator. 
However, one finds that there is an imaginary part associated with the (localized) Hamiltonian when there is particle flux through the domain boundary,see~\cite{Kirsch:1981sf}, and extended discussion in Sections 9.1 and 9.5 in Ref.\,\cite{GMR85}
\begin{equation}
\hat H=\dsfrac{1}{2}\int d^3x[\hat\Psi^\dagger(\vec x,t),H_D\hat\Psi(\vec x,t)]
+\dsfrac{i}{2}\oint d\vec\sigma\cdot[\hat\Psi^\dagger,\vec\alpha\hat\Psi]
\;.\label{eq:eqO-4-8}
\end{equation}

In the same way one can construct an operator for the charge-current density:
\begin{equation}
\hat j^\mu(\vec x)=\dsfrac{e}{2}[\hat\Psi^\dagger(\vec x),\gamma^0\gamma^\mu\hat\Psi(\vec x)]\;, 
\label{eq:eqO-4-9}
\end{equation}
and find the total charge operator:
\begin{equation}
\hat Q=\dsfrac{e}{2}\int d^3x[\hat\Psi^\dagger,\hat\Psi]
\;.\label{eq:eqO-4-10}
\end{equation}
By explicit calculation it is easy to show that $\hat{Q}$ is a constant of motion (except for surface effects):
\begin{equation}
\dsfrac{d\hat Q}{dt}=i[\hat H,\hat Q]=-\oint\vec\sigma\cdot\hat{\vec j}
\;.\label{eq:eqO-4-11}
\end{equation}
This equation allows us to rewrite \req{eq:eqO-4-8} in the following way:
\begin{equation}
\label{eq:eqO-4-12}
\hat H=\hat H_\mathrm{loc}+\dsfrac{i}{e}\dsfrac{d}{dt}\hat Q\;,\qquad 
\hat H_\mathrm{loc}\equiv\dsfrac{1}{2}\int d^3x[\hat\Psi^\dagger,H_D\hat\Psi]\;.
\end{equation}
Let us for the moment neglect surface effects. In the representation of single-particle states the Hamiltonian and the charge operator take the following form:
\begin{equation}
\hat H_\mathrm{loc}=\frac 1 2 \sum_{n> F}E_n[\hat b^\dagger_n,\hat b_n]+
\frac 1 2 \sum_{n< F}(-E_n)[\hat d^\dagger_n,\hat d_n]\;,
\label{eq:eqO-4-13}
\end{equation}
and we find for the vacuum energy 
\begin{equation}
E_\mathrm{V}\equiv \langle \mathrm{V}|\hat H_\mathrm{loc}|\mathrm{V}\rangle =-\frac{1}{2}\sum_{n}^{}|E_n|< 0 \;.\label{eq:eqO-4-13A}
\end{equation}
Similarly we obtain for the the vacuum charge
\begin{equation}
\hat Q=\frac e 2\left( \sum_{n> F}[\hat b^\dagger_n,\hat b_n]
-\sum_{n< F}[\hat d^\dagger_n,\hat d_n] \right)\;,
\label{eq:eqO-4-14}
\end{equation}
with
\begin{equation}
Q_\mathrm{V}\equiv \langle \mathrm{V}|\hat Q|\mathrm{V}\rangle =-\dsfrac{e}{2}\left(\sum_{n> {\rm F}}1- \sum_{n< {\rm F}}1\right)\;. \label{eq:eqO-4-14A}
\end{equation}
For given external potential and in particular for a potential that we follow as it turns from subcritical to supercritical, $E_\mathrm{V}^\mathrm{loc}$ \req{eq:eqO-4-13A} changes smoothly across the critical point and more generally, it does not depend on the detail of the vacuum state properties as it is summed up over all single-particle eigen energies. Therefore we do not need to follow up this quantity in the present context. On the other hand the vacuum charge $Q_\mathrm{V}$ evaluates the difference in the single-particle counts of Hilbert space sectors of particles and antiparticles. This quantity will exhibit a major change when the external potential alters the count of states in these two sectors of the Hilbert space.

\subsection{Supercritical vacuum state}\label{Sec32}
\subsubsection{Weak field limit}\label{Sec321}
The state of lowest energy,
\ie\ with the lowest expectation value of $\hat{H}_\mathrm{loc}$, is the one that
is an eigenstate of eigenvalue zero with respect to all operators
$\hat{N}^e_n$ and $\hat{N}^p_n$, in combination with the following choice of the Fermi
surface $\mathrm{F}_0$ which we shall also denote by $E_\mathrm{F}=0$:
\begin{equation}
E_n> 0\ :\ n> \mathrm{F}_0\ ,\ \ \ E_n< 0\ :\ n< \mathrm{F}_0
\;.\label{eq:eqO-4-15}
\end{equation}

We obtain the state of lowest energy by dividing electron
and positron states according to the sign of the energy eigenvalue
and requiring that no particle or antiparticle be present.
We shall call this state \emph{the absolute ground state} or \emph{state of
lowest energy} $|0,\mathrm{F}_0\rangle $:
\begin{align}\label{eq:eqO-4-16}
\hat b_n|0,\mathrm{F}_0\rangle &=0\ :\ n> \mathrm{F}_0 \;, \\ \nonumber
\hat d_n|0,\mathrm{F}_0\rangle &=0\ :\ n< \mathrm{F}_0
\;.
\end{align}

For a vanishing external potential the Dirac equation is charge
conjugation invariant, and we have an equal number of states with
$n> \mathrm{F}_0$ and $n< \mathrm{F}_0$. As a consequence the ground state will have zero charge:
\begin{equation}
Q_\mathrm{V}(\mathrm{F}_0)=\langle 0,\mathrm{F}_0|\hat Q |0,\mathrm{F}_0\rangle =0
\;.\label{eq:eqO-4-17}
\end{equation}

\subsubsection{Supercritical fields}\label{Sec322}
Consider an external attractive potential for electrons with a strength parameter $\lambda$
\begin{equation}
V_\lambda(\vec x)=\lambda v(\vec x)
\;.\label{eq:eqO-4-18}
\end{equation}
According to the discussion of Section~\ref{Sec2}, for the same strength $\lambda_1$ the most strongly bound-state acquires a binding energy equal to the rest mass $m$ of the electron. For $\lambda> \lambda_1\ :\ E(\lambda)< 0$ this level is counted as a positron state. Therefore it is shifted from the sum over $n> \mathrm{F}_0$ to the sum $n< \mathrm{F}_0$.

This changes the balance in the expression for $Q_\mathrm{V}$:
\begin{equation}
\langle 0,\mathrm{F}_0|\hat Q |0,\mathrm{F}_0\rangle =Q_\mathrm{V}(\mathrm{F}_0)=
eN(\lambda)\Theta(\lambda-\lambda_1)
\;.\label{eq:eqO-4-19}
\end{equation}
where $N(\lambda)$ denotes the number of states with a binding energy exceeding the rest mass $m$. We conclude that beyond a certain strength of the external potential the lowest energy state of the electron- positron field carries a none zero charge.

This state can only be reached if precisely the required number of electrons is supplied. Interesting as it may be, the lowest energy state is therefore a purely formal construction since the charge operator $\hat{Q}$ is a constant of motion according to \req{eq:eqO-4-11} as long as surface effects can be neglected. When the binding energy of a bound-state is increased too much beyond $m$, its wave function remains localized - the surface effects vanish. The situation is fundamentally different when the strength of the external potential is increased to the point $\lambda_\mathrm{cr}$ where one of the bound states is bound by twice the electron rest mass, $2m$. 

As we discussed in Section~\ref{Sec2}, for $\lambda> \lambda_\mathrm{cr}$ the bound-state becomes embedded into the antiparticle scattering states as a resonance state. According to \req{eq:eqO-4-11} the localized charge of the vacuum state can change as particles (or antiparticles) cross the boundary and at the same time the local Hamiltonian $\hat{H}_\mathrm{loc}$ acquires an imaginary part indicative of a decay process. In particular the possibility of exchanging particles with the surrounding infinity develops. 

All this means that the vacuum can make a transition from one charge subspace $\mathbb{V}_q$ of the total Fock space to another subspace $\mathbb{V}_{q\prime }$ by the emission of an antiparticle (or particle). Each subspace is characterized by a different eigenvalue of the charge operator. In each sector (subspace) of the Fock space there is a state of lowest energy, the equilibrium state. It is most easily determined as the state that minimizes
\begin{equation}
\hat I=\hat H_\mathrm{loc}+\mu\hat Q\;,
\label{eq:eqO-4-20}
\end{equation}
where it can be shown that the quantity $\mu$, the chemical potential,
must be chosen as
\begin{equation}
\mu=\frac{m}{e}\;,\nonumber
\end{equation}
in order to ensure that pair production is responsible for a transition from one charge sector to another, one member of the pair being emitted to infinity. We thus find the following condition for the equilibrium state:
\begin{equation}
\langle \mathrm{equil}|\hat{H}_\mathrm{loc}+\frac{m}{e}\hat{Q}|\mathrm{equil}\rangle =\mathrm{min} \;.\nonumber
\end{equation}
By means of Eqs. \ref{eq:eqO-4-13},\ \ref{eq:eqO-4-14} we can rewrite the operator $\hat{H}_\mathrm{loc}+\frac{m}{e}\hat{Q}$ as:
\begin{align}\label{eq:eqO-4-21}
\hat{H}_\mathrm{loc}+\frac{m}{e}\hat{Q}= 
&\sum_{n> F}(E_n+m)\hat b^\dagger_n\hat b_n+
\sum_{n< F}(-E_n-m)\hat d^\dagger_n\hat d_n \\ \nonumber
&+(E_\mathrm{V}+\dsfrac{m}{e}Q_\mathrm{V})
\;.
\end{align}
Following the above line of arguments the equilibrium state with the lowest expectation value of $\hat{I}$ is found by requiring 
\begin{align}\label{eq:eqO-4-22}
\hat b_n|\mathrm{equil}\rangle &=0 \ :\ n> \mathrm{F}_{-m}\;, \\ \nonumber
\hat d_n|\mathrm{equil}\rangle &=0 \ :\ n< \mathrm{F}_{-m}
\;,
\end{align}
where the Fermi surface is chosen according to
\begin{equation}
E_n+m> 0\ :\ n> \mathrm{F}_{-m}\ ,\ \ \ E_n+m< 0\ :\ n< \mathrm{F}_{-m}
\;,\label{eq:eqO-4-23}
\end{equation}
\ie\ the Fermi energy is $E_\mathrm{F}=-m$. The state, $|\mathrm{equil}> =|0,\mathrm{F}_{-m}\rangle $ is the state of an atomic system subject to a given external potential in the absence of interference from outside. In this state, all levels with $E> -m$ are particle states and all levels with $E< -m$ are antiparticle states. It is precisely the state we have called the \emph{charged vacuum state} (for $\lambda> \lambda_\mathrm{cr}$) in Section~\ref{Sec2}. We have now shown that a neutral atomic system in a weak external field will become the state $|0,\mathrm{F}_{-m}\rangle $ after the potential has been increased to arbitrary strength and sufficient time has elapsed for the equilibrium to be established.

Let us summarize our results. There are two different possible definitions of the vacuum state:

1. The state of absolutely lowest energy $|0\rangle $, which is characterized by the Fermi energy $E_\mathrm{F}=0$, \req{eq:eqO-4-16}. Particle and antiparticle states are divided according to the sign of the energy eigenvalue. Due to conservation of electric charge, a microscopic system can often not reach this state.

2. In practice, the system will change its charge by antiparticle (or particle) emission until it reaches the \lq\lq charged vacuum\rq\rq\ equilibrium state $|0,Q_0\rangle $, which is characterized by the Fermi energy $E_\mathrm{F}=-m$, \req{eq:eqO-4-22} . All (also the former bound) states below $E=-m$ are counted as antiparticle states. Whenever sufficient time is available, any system will spontaneously occupy this state. Of these two definitions, the charged vacuum is therefore the one with the greatest importance.

\subsubsection{Propagators in supercritical fields}\label{Sec323}
We now reconsider these results from the point of view of the Green\rq s function~\cite{[66]}. As before we focus here on the case of a particle moving in a time-independent potential $A_\mu$. The Green\rq s function satisfies the equation
\begin{equation}
(i\gamma^\mu\partial_\mu-e\gamma^\mu A_\mu-m)G(x,x^\prime)=\delta^4(x-x^\prime)
\;.\label{eq:eqO-4-25}
\end{equation}
Because of the time independence of the potential, the Green\rq s function must be invariant under displacements in time. Thus, the Green\rq s function may be represented as the Fourier transform
\begin{equation}
G_C(x,x^\prime)=\int_C\dsfrac{d\omega}{2\pi}e^{-i\omega(t-t^\prime)}G(\vec x,\vec x^\prime;\omega)
\;.\label{eq:eqO-4-26}
\end{equation}
The choice of the contour $C$ is related to boundary conditions satisfied by $G(x,x\prime)$ as $t\rightarrow\pm \infty$. It plays the same role as the choice of the Fermi energy $E_\mathrm{F}=-m$, \req{eq:eqO-4-23}, in the Hamiltonian approach, which makes the distinction between particles and anti-particles. The conventional choice of $C$, which leads to the Feynman-St\"uckelberg boundary conditions, is shown in \rf{Fig_m07}. There, the two branch cuts beginning at $\omega=\pm m$ as well as the poles associated with the bound states are shown. The integrand of \req{eq:eqO-4-26} may be represented as a sum over the entire spectrum of eigensolutions of Dirac equation, namely
\begin{equation}
G(\vec x,\vec x^\prime;\omega)=\sum_E\dsfrac{\Psi_E(\vec x)\bar\Psi_E(\vec x^\prime)}{\omega-E}
\;.\label{eq:eqO-4-27}
\end{equation}
\begin{figure}[htb]
\centering{
\resizebox{0.95\textwidth}{!}{%
	\includegraphics{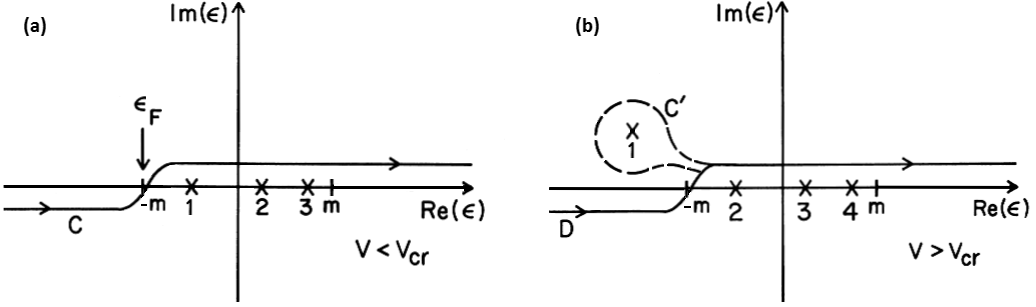}
}
}
\caption[]{a) The conventional choice of the contour in the complex $\omega$ plane for sub-critical fields $V<V_\mathrm{cr}$. The contour $C$ crosses the real axis at the Fermi energy. b) Two choices of contours, $C\rq$ and $D$ for $V>V_\mathrm{cr}$. The  contour $D$ corresponds to a stable charged vacuum state.}\label{Fig_m07}
\end{figure}
Substituting this expression into \req{eq:eqO-4-26} and using the contour of \rf{Fig_m07}a leads to the representation
\begin{align}\label{eq:eqO-4-28}
G(x,x^\prime)=&-i\Theta(t-t^\prime)\sum_{E> E_\mathrm{F} }\Psi_E(\vec x)\bar\Psi_E(\vec x^\prime)e^{-iE(t-t^\prime)}
\\ \nonumber
&+i\Theta(t-t^\prime)\sum_{E< E_\mathrm{F} }\Psi_E(\vec x)\bar\Psi_E(\vec x^\prime)e^{-iE(t-t^\prime)}
\;.
\end{align}
Our treatment of the Green\rq s function is so far identical to the usual discussions of the Feynman propagator, which may be conveniently defined as
\begin{equation}
S_F(x,x^\prime)=-i\langle 0|T(\hat\Psi(x),\hat{\bar\Psi}(x^\prime))|0\rangle 
\;,\label{eq:eqO-4-29}
\end{equation}
where $T$ denotes the time-ordered product. Substituting the expansions of $\hat{\Psi}(x)$ and $\hat{\bar{\Psi}}(x^\prime )$ in terms of the solutions of the Dirac equation leads to the expression on the right-hand side of \req{eq:eqO-4-28} for $S(x,x\rq)$, which establishes that $S$ and $G$ are identical in the case of weak fields, provided that one makes the distinction between particles and antiparticles that is consistent with the contour of \rf{Fig_m07}a.

As discussed above in the supercritical case, it is important for the Fermi energy to remain at $E_\mathrm{F}=-m$ in order to have a stable state of reference. The behavior of $G(\vec{x},\vec{x} \prime ;\omega)$ in the complex $\omega$-plane leads to a similar criterion: as the potential strength is increased from an subcritical value to an supercritical value, the pole associated with the lowest bound-state in \rf{Fig_m07}a moves off the real axis and into the upper half of the complex plane as shown in \rf{Fig_m07}b. 

It is important to appreciate that this singularity is on the second sheet~\cite{[65]} and that the contour $C$ is not deformed (into the contour $C\rq$) so as to continue to embrace the pole. Instead it is necessary to choose the contour $D$, where the Fermi energy remains at $-m$. The path $C\rq$ corresponds to the choice of the neutral vacuum as the reference state, which is not stable. 

Now we show that the choice of contour $D$ leads to a reasonable result and that the choice of contour $C\rq$ does not. Substituting the first term of \req{eq:eqO-2-21} into \req{eq:eqO-4-27} for $G$, we obtain
\begin{equation}
G_D(\vec x,\vec x^\prime;\omega)\sim
\int_{-\infty}^{\infty}dE\dsfrac{|a(E)|^2}{\omega-E-i\eta}\Psi^{cr}(\vec x)\bar\Psi^{cr}(\vec x^\prime)\;,
\label{eq:eqO-4-30}
\end{equation}
where only the interesting part of $G$ has been kept and $\eta$ is negative when $\omega< -m$ and positive when $\omega> -m$, as required by the choice of contour $D$. From \req{eq:eqO-4-22} it is apparent that $a(E)$ carries the singularity associated with the resonance, that is, the pole shown in the upper half plane of \rf{Fig_m07}b. This pole, however, occurs on the second sheet and the only contribution to the integral of \req{eq:eqO-4-30} arises from the pole at $E=\omega-i\eta$ (provided that $\omega> -m$).

Thus the result of the integration is
\begin{equation}
G_D(\vec x,\vec x^\prime;\omega)\sim i\dsfrac{\Gamma\Theta(-m-\omega)}{(\omega-E_{res})^2+\Gamma^2/4}\Psi^{cr}_0(\vec x)\bar\Psi^{cr}_0(\vec x^\prime)\;,
\label{eq:eqO-4-31}
\end{equation}
where we have treated the resonance approximately as discussed in Section~\ref{Sec2}. A very different result would have been obtained if we had chosen the contour $C\rq$. Then the pole at $E=E_{res}+i\Gamma /2$ makes a contribution of the form
\begin{equation}
G_{C^\prime}(\vec x,\vec x^\prime;\omega)\sim\dsfrac{\Psi_0^{cr}(\vec x)\bar\Psi_0^{cr}(\vec x^\prime)}{\omega-E_{res}-i\Gamma/2}\;,
\label{eq:eqO-4-32}
\end{equation}
which is characteristic for a complex eigenvalue, a reflection of the lack of stability of the state of reference defined by the choice $C\rq$. 

To summarize, every time a bound-state descends into the negative energy continuum, we must redefine the Green\rq s function so as to include only the remaining poles on the real axis. This is done by maintaining the fixed shape $D$ of the contour. As described this implies a change in the charge of the vacuum each time a pole crosses the fixed integration path $D$ in \rf{Fig_m07}.

\subsection{QED and supercritical fields}\label{Sec33}
\subsubsection{Self consistent equations for single-particle states}\label{Sec331}
In QED of strong fields one has to deal with two different parameters defining the coupling strength, namely $\alpha$ and $Z\alpha$. In the heaviest stable elements $Z\alpha\simeq0.7 $ and it can exceed unity in superheavy (quasi-molecular) systems, see below Section~\ref{Sec41}. Thus the usual series expansion in $(Z\alpha)^n\alpha^m$ becomes questionable. We will therefore describe a method of evaluating the usual QED corrections based on the exact Dirac propagator in the external Coulomb field, and in doing this we will include all terms $(Z\alpha)^n$. In this procedure  the radiation field effects characterized by the small constant $\alpha$ can then be treated as a perturbation.

The ground state expectation value of the current operator is
\begin{equation}
\langle 0|\hat j_\mu|0\rangle =\mathrm{Tr}(iS_{\mathrm{F}}(x,x)\gamma_\mu)=
\widetilde{\sum_q}\bar\Phi_q\gamma_\mu\Phi_q\;,
\label{eq:eqO-6-9}
\end{equation}
where the propagator at the point $x = y$ is defined by the
prescription
\begin{equation}
S_{\mathrm{F}}(x,x)=\dsfrac{1}{2}\lim_{\epsilon\to 0}[S_{\mathrm{F}}(x,x+\epsilon)+S_{\mathrm{F}}(x,x-\epsilon)]
\;,\label{eq:eqO-6-10}
\end{equation}
with $\epsilon$ time-like, and the \lq\lq tilde sum\lq\lq 
\begin{equation}
\widetilde{\sum_q}=\dsfrac{1}{2}\left( \sum_{E_q> E_{\mathrm{F}}}-\sum_{E_q< E_{\mathrm{F}}}\right)=\sum^{HF}_{q}+\dsfrac{1}{2}\left( \sum_{E_q< -m}-\sum_{E_q> -m}\right)
\;.\label{eq:eqO-6-11}
\end{equation}
It equals the Hartree-Fock sum $\sum\limits_{q}^{HF} = \sum\limits_{-m< q< E_\mathrm{F}}^{}$ over all occupied bound states plus the \lq\lq Uehling sum\lq\lq. The latter usually accounts for the effects of (virtual) vacuum polarization since it describes the induced current due to the presence of the external source. 

Relation \ref{eq:eqO-6-9} is the basic starting point~\cite{[66]} for the calculation of the vacuum polarization in strong external fields exactly in all orders of the external field. In order to derive a system of classical self-consistent one-particle equations for the set $\Phi_q$, one considers the matrix elements of the equation of motion between the vacuum and the single-particle (or hole) state $\hat{b}_q^\dagger|0\rangle $, $\hat{d}_q^\dagger|0\rangle $. Using the definition of the current operator \req{eq:eqO-4-9} and the commutation relations of the field operators the following set of equations is obtained:
\begin{subequations}
\begin{align}\label{eq:eqO-6-12}
(-i\vec\alpha\cdot\vec\nabla+V_{ex}+\beta m-\epsilon_n)\Phi_n(\vec x) &=
\int d^3z\dsfrac{e^2}{4\pi}\left(\dsfrac{\widetilde\sum_{n^\prime}\bar \Phi_{n^\prime}(\vec z)\gamma_\mu\Phi_{n^\prime}(\vec z)}{|\vec x-\vec z|}\right)\beta\gamma^\mu\Phi_{n}(\vec x)
-\\ \nonumber
&\hspace*{-1cm}- \int d^3z\dsfrac{e^2}{4\pi}\widetilde\sum_{n^\prime}\dsfrac{\bar\Phi_{n}(\vec z)\gamma_\mu \Phi_{n^\prime}(\vec z)F_{n^\prime n}(|\vec x-\vec z|)}{|\vec x-\vec z|}\beta\gamma^\mu\Phi_{n^\prime}(\vec x)
\;,
\end{align}
where
\begin{equation}
F_{n^\prime n}(|\vec x-\vec z|)=-\dsfrac{2}{\pi}\int_0^\infty dq\dsfrac{\sin(q|\vec x-\vec z|)}{\epsilon_{n^\prime}-\epsilon_n+\sigma q}
\;.\label{eq:eqO-6-13}
\end{equation}
\end{subequations}
In this equation $\sigma= + 1$ for $\epsilon> \epsilon_\mathrm{F}$ and $-1$ for 
$\epsilon< \epsilon_\mathrm{F}$.

\req{eq:eqO-6-12} was derived~\cite{[68],[67b],[69]} from the Schwinger-Dyson equations for the electron propagator. This self-consistent system of equations is obtained by neglecting radiative corrections to the vertex function and the photon propagator, a procedure that is equivalent to the Hartree-Fock approximation in the non-relativistic many-body theory. As long as the potential is subcritical, it is simple to isolate the various contributions to \req{eq:eqO-6-12} and to determine the charge of the system. 

Let us consider a superheavy nucleus surrounded by enough atomic electrons such that the atom is neutral. Then the last term of first line in \req{eq:eqO-6-12} is, according to \req{eq:eqO-6-11}, the sum of the direct term of the Hartree-Fock approximation as well as the vacuum polarization correction. The last term (2nd line of \req{eq:eqO-6-12}) includes the exchange term of the Hartree-Fock approximation as well as the electromagnetic self-energy corrections. The retardation correction, which the function $F$ represents, is usually neglected in the Hartree-Fock calculations. Both the vacuum polarization and self-energy terms in \req{eq:eqO-6-12} have to be renormalized to recover the physically observable quantities.

The self-consistent solution of \req{eq:eqO-6-12} is a formidable task. Fortunately, the smallness of the fine structure constant never allows the field corrections to become very large in the subcritical case. Restricting the self-consistent approach to the electrons in the bound states is an adequate first step.

In the supercritical case, where vacuum polarization effects become {\em real}, the self-consistent treatment of field effects may become important and \req{eq:eqO-6-12} furnishes an acceptable starting point. At the end of this section we discuss a self-consistent treatment~\cite{[70]} of the real vacuum polarization screening that uses the relativistic Thomas-Fermi (RTF) model to describe the many-body effects. The self-consistent method provides a means for justifying the RTF approach, but a simpler derivation will be presented below.

\subsubsection{Virtual vacuum polarization effects}\label{Sec332}
We now describe the effects of virtual vacuum polarization in high $Z$ atoms. The vacuum polarization charge density and its corresponding potential produced by the nuclear Coulomb field was calculated to first order in $Z\alpha$ at first for a point-like nucleus by Serber~\cite{[71]}, and by Uehling~\cite{[72]} 
\begin{equation}
V_{\rm VP}(r)=-\dsfrac{2\alpha Z\alpha}{3\pi r}\int_m^\infty d\tau e^{-2\tau r}(1+\dsfrac{m^2}{2\tau^2})(\tau^2-m^2)^{1/2}\tau^{-2}
\;.\label{eq:eqO-6-14}
\end{equation}
It is easily seen from \req{eq:eqO-6-14} that $V_{\rm VP}$ vanishes exponentially for $r\gg 1/m$. Therefore vacuum polarization has only an extremely small influence on wave functions that have an extension large compared with the electron Compton wavelength. For example, the energy shift due to vacuum polarization in the Lamb-shift of hydrogen is only -27 MHz (compared with +1079 MHz from the self-energy correction). However, it increases strongly if the wave function becomes more localized. This is in particular true for muonic atoms where it is the dominant QED correction.

In one of the heaviest atomic systems accessible for spectroscopy, the element Fermium Fm (Z=100), vacuum polarization produces an energy shift of 155 eV out of 142 keV total binding energy for the $1s_{1/2}$ state~\cite{[73]}. At even higher $Z$ Uehling energy-shift was calculated by Pieper and Greiner~\cite{[7]}. It approximately doubles if $Z$ is increased by 10 units of charge due to the collapse of the electrons to the nucleus and reaches $ E =- 11.83$ keV for the $1s_{1/2}$ state at $Z=171$. We note that it increases the binding slightly and so makes $Z_\mathrm{cr}$ smaller by about 1/3 of a proton charge (but compensated largely by self-energy effects, see below).

Explicit calculations have confirmed that higher-order contributions remain small even for $Z\alpha > 1$. Wichmann and Kroll~\cite{[66]}  were the first to develop a method to calculate vacuum polarization to all orders $\alpha (Z\alpha)^n$ employing the exact (single-particle) solutions of the Dirac equation in the external field. The method was later applied by Rinker and Wilets~\cite{[74a],[74b]} and by Gyulassy~\cite{[75a],[75b],[75c]} to treat also extended nuclei with $Z\alpha > 1$ including supercritical nuclei. With a nuclear radius R = 10 fm the critical charge is $Z_\mathrm{cr}\alpha= 1.27459$. At $Z\alpha= 1.27445$ the $1s_{1/2}$ energy is just above the negative continuum, namely $E_{1s_{1/2}} = -.999$. Here the energy shift~\cite{[75a],[75b],[75c]} due to vacuum polarization is $\Delta E^3 = 0.570$keV and $\Delta E^{3+} = 1.150$keV. This demonstrates, that the Uehling potential leads by far to the strongest energy shift ($\Delta E^1$) and higher orders do not qualitatively change the behavior of the diving bound-state.

Using a monopole approximation to simulate the U+U quasi-molecule near the diving point Rinker and Wilets~\cite{[74a],[74b]} found an energy shift of -3.98keV consisting of $\Delta E^1 = -4.62$keV from the Uehling potential $\Delta E_1^{3+} = +609$eV, $\Delta E_2^{3+} = +34$eV for $|\kappa| = 1,2$. As we have discussed at length before, at $Z> Z_\mathrm{cr}$ the pole corresponding to the $1s_{1/2}$ state moves off the physical sheet for the Green\rq s function G. This necessitates the introduction of a charged vacuum since the contour $C$ is not able to follow the $1s_{1/2}$ pole and has to remain inside the gap between $-m$ and $+m$. As the potential strength is increased from an subcritical value to an supercritical value, the vacuum polarization charge density changes discontinuously.

\subsubsection{Real vacuum polarization}\label{Sec333}
It is important to realize that for $Z> Z_\mathrm{cr}$ the vacuum polarization can be broken up into two terms 
\begin{equation}
\rho_{VP}(\vec x)=\rho_{VP}^V(\vec x)+\rho_{VP}^R(\vec x)\;,
\label{eq:eqO-6-21}
\end{equation}
the first of which (the \lq\lq virtual\lq\lq  vacuum polarization) is a smooth extrapolation of the vacuum polarization charge density for $Z< Z_\mathrm{cr}$, whereas the second term (the so-called \lq\lq real\lq\lq  vacuum polarization) goes over continuously into the charge distribution of the bound-state just before diving occurs. As far as the vacuum is concerned, all of the effects of the discontinuity are included in the real part. 

As we have described above, the resonance in the lower continuum gives rise to a singularity on the second sheet in the complex $\omega$-plane, see \rf{Fig_m07}b. The real vacuum polarization charge density may in principle be calculated by
\begin{equation}
\rho_{VP}=-ie\int_C\dsfrac{d\omega}{2\pi}\mathrm{Tr}[\gamma^0G(\vec x,\vec x^\prime;\omega)]_{\vec x^\prime\to \vec x}\;,
\label{eq:eqO-6-16}
\end{equation}
where the full contour $C$ is replaced. The closed contour $R$ surrounding the singularity on the second sheet is shown in\rf{Fig_m08}.
\begin{figure}[t]
\centering{
\resizebox{0.5\textwidth}{!}{%
\includegraphics{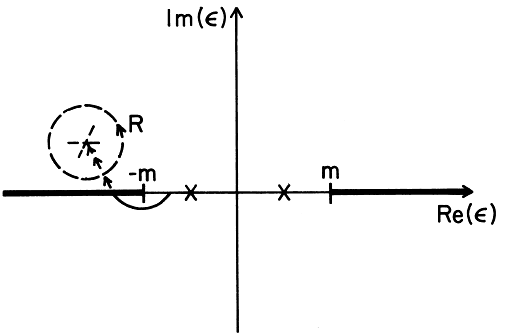}
}
}
\caption[]{The contour $R$ surrounding the pole on the
second sheet.}\label{Fig_m08}
\end{figure}

The origin of this contribution is simply the change in the charge density arising from the change in the definition of the Green\rq s function when the ls orbit becomes supercritical. Thus $\rho_{VP}^R(\vec{x})$ is intimately related to the residue of the Green\rq s function at the pole on the second sheet. It is easy to verify the consistency of this definition for the approximate treatment of the contribution of the resonance to the Green\rq s function carried out above. Inserting \req{eq:eqO-4-31} into \req{eq:eqO-6-16} one finds that
\begin{equation}
\rho_{VP}^R(\vec x)\approx e\Psi_0^{cr\dagger}(\vec x)\Psi_0^{cr}(\vec x)\;,
\label{eq:eqO-6-22}
\end{equation}
after carrying out the $\omega$ integration. This shows that
$\rho_{VP}^R$ is just the smooth continuation of the bound-state
charge density.

The real vacuum polarization density for several supercritical potentials is shown in \rf{Fig_m09} ~\cite{[23]}. It is interesting to compare the results for $Z = 172$ and $Z = 184$. The result for $Z = 184$ suggests that the real vacuum polarization charge density continues to shrink as the nuclear charge is increased. The calculations were carried out approximately. The exact expression for the s state contribution in the supercritical basis leads to
\begin{align}\label{eq:eqO-6-23}
\rho_V(\vec x)&=\dsfrac{e}{2}\int_{-\infty}^{-m}d\epsilon\Psi_\epsilon^\dagger(\vec x)\Psi_\epsilon(\vec x)
-\dsfrac{e}{2}\int_{m}^{\infty}d\epsilon\Psi_\epsilon^\dagger(\vec x)\Psi_\epsilon(\vec x) 
\\ \nonumber & \ \ \ - \dsfrac{e}{2}\sum_{\epsilon_\beta\neq1s}\Psi^\dagger_{\epsilon_\beta}(\vec x)\Psi_{\epsilon_\beta}(\vec x)
\;,
\end{align}
where the first term includes the effects of the resonance. Initially, an energy interval centered on the resonance was chosen and
\begin{equation}
\rho_V^R(\vec x)\approx\dsfrac{e}{2}\int_{\epsilon_-}^{\epsilon_+}d\epsilon[\Psi_\epsilon^\dagger(\vec x)\Psi_\epsilon(\vec x)-\Psi_\epsilon^\dagger(\vec x)\Psi_{-\epsilon}(\vec x)]
\;,\label{eq:eqO-6-24}
\end{equation}
was computed where $\epsilon_\pm =\epsilon \pm 5\Gamma$, thus incorporating the symmetry between positive and negative values of $\epsilon$. A better method~\cite{[23]} for isolating the contribution of the real part is based on
\begin{equation}
\rho_V^R( x)=2(\rho_V(x)-\rho_{V-\delta V}(x))\;,\label{eq:eqO-6-25}
\end{equation}
where $\delta V$ is chosen such that the potential $V- \delta V$ does not generate a resonance in the interval $(\epsilon_-,\epsilon_+)$. Thus, the second term amounts to a subtraction of the effects of virtual vacuum polarization. Gyulassy~\cite{[75a],[75b],[75c]} also calculated the charged densities of the supercritical vacuum using the connection between the Green\rq s function and the charge density. His results agreed with those of the Frankfurt group and confirmed that the size of the region occupied by $\rho_{VP}^R(\vec{x})$ continues to shrink as $Z$ is increased beyond $Z_\mathrm{cr}$. 
\begin{figure}[t]
\centering{
\resizebox{0.7\textwidth}{!}{%
\includegraphics{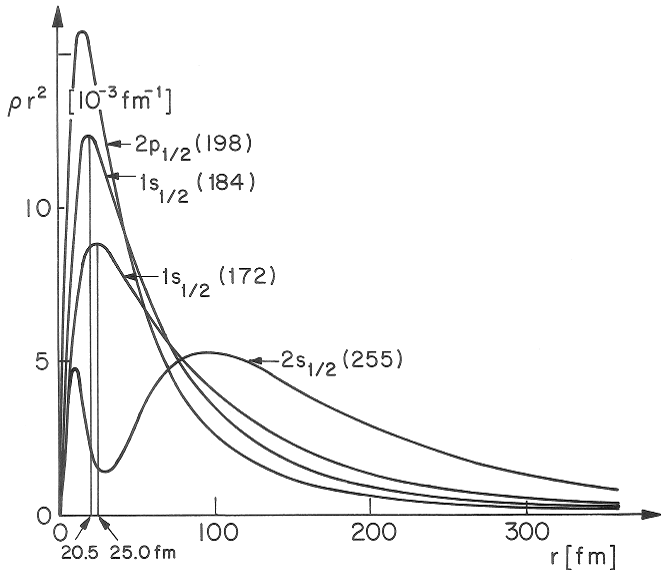}
}
}
\caption[]{Real vacuum polarization charge densities for
several supercritical potentials, Ref.\,\cite{[12]}.}\label{Fig_m09}
\end{figure}

We now turn to consider the screening effect of the \lq real\rq\ vacuum polarization. We recall that the point at which the $1s$-wave function joins the continuum solutions of negative frequency has been determined to be about $Z = 173$, under \lq\lq realistic\rq\rq\  assumptions and extrapolations of the known properties of nuclear and electromagnetic interactions. Similarly, the next critical point at which the $2p_{1/2} $ state is expected to join the continuum is about $Z = 185$. At this point the charge of the vacuum increases to $4e$. Soon, as we increase the nuclear charge, higher angular momentum states will also join the lower continuum, and the charge of the vacuum will rise even faster. Thereafter, the accumulating {\it negative} charge of the electron vacuum will increasingly screen the attractive force of the positive nuclear charge requiring an explicit description of the {\it back reaction} of the charged vacuum onto itself.

It has been proposed~\cite{[70]} to make use of the relativistic Thomas-Fermi approximation for sufficiently large $Z-Z_\mathrm{cr}$, when many states have joined in the lower continuum. The charge density of the vacuum is equal to the charge density carried by all the states that have joined the lower continuum. In the Thomas- Fermi model, the sum over all these states is represented by an integral over all states with momentum inside the Fermi sphere of radius $k_\mathrm{F}$. The density of electrons is related to the Fermi momentum $k_\mathrm{F}(x)$ by
\begin{equation}
\rho_e=\dsfrac{e}{3\pi^2}k_\mathrm{F}^3
\;. \label{eq:eqO-9-1}
\end{equation}
The effect of the spin degeneracy is included in \req{eq:eqO-9-1}. The relativistic relation between the Fermi energy $E_\mathrm{F}$ and Fermi momentum is 
\begin{equation}
k_\mathrm{F}^2=[(E_\mathrm{F}-eV)^2-m^2]\Theta(E_\mathrm{F}-eV-m)
\;. \label{eq:eqO-9-2}
\end{equation}
The step function ensures that $k_\mathrm{F}^2$ is a positive quantity.
From \req{eq:eqO-9-1} and \req{eq:eqO-9-2} we now obtain for the charge density
of the ground state $|\Omega\rangle $ characterized by a choice of $E_\mathrm{F}$:
\begin{equation}
\langle \Omega|\rho_e|\Omega\rangle =\dsfrac{e}{3\pi^2}[(E_\mathrm{F}-eV)^2-m^2]^{3/2}\Theta(E_\mathrm{F}-eV-m)
\;. \label{eq:eqO-9-3}
\end{equation}
Introducing the total charge density $\rho_T$ which is composed of the
external \lq\lq nuclear\rq\rq\  part $\rho_N$ and the electronic part
\begin{equation}
\rho_T=\rho_N+\langle \Omega|\rho_e|\Omega\rangle 
\;, \label{eq:eqO-9-4}
\end{equation}
and using Coulomb\rq s law
\begin{equation}
\Delta eV(\vec r)=-e\rho_T(\vec r)
\;, \label{eq:eqO-9-5}
\end{equation}
we find a self-consistent non-linear differential equation for the average potential $V$, that depends on the choice of the Fermi surface $E_\mathrm{F}$ characterizing the ground state
\begin{equation}
\Delta eV(\vec r)=-e\rho_N(\vec r)-\dsfrac{e^2}{3\pi^2}[(E_\mathrm{F}-eV)^2-m^2]^{3/2}\Theta(E_\mathrm{F}-eV-m)
\;. \label{eq:eqO-9-6}
\end{equation}
As long as the nuclear background charge $\rho_N$ is isolated from external sources of electrons, the proper choice of $E_\mathrm{F}$ is $E_\mathrm{F}=-m$, \req{eq:eqO-4-22}. If this condition is relaxed and an inexhaustible supply of electrons is available, we must account for only the kinetic energy of these electrons. Thus for neutral atomic system we must take $E_\mathrm{F}=m$, which furthermore gives in the limit $|-2mV|> |V^2|$ the usual nonrelativistic Thomas-Fermi model. 

We now consider \req{eq:eqO-9-6} with the Fermi energy fixed at $E_\mathrm{F}=-m$. This means that only the states accessible to spontaneous decay are filled. Inserting $E_\mathrm{F}=-m$ into \req{eq:eqO-9-6} yields
 \begin{equation}
\Delta eV(\vec r)=-e\rho_N(\vec r)-\dsfrac{e^2}{3\pi^2}(2meV+e^2V^2)^{3/2}\Theta(-eV-2m)
\;. \label{eq:eqO-9-7}
\end{equation}

We now proceed to discuss the solution of \req{eq:eqO-9-7}. Since the charge density of the vacuum must be confined to the vicinity of the external charge, we require a solution such that
\begin{equation}
 eV(r) \rightarrow \ -\frac{\gamma \alpha}{r}\ \ \ \mbox{for}\ r\ \rightarrow\ \infty\;, \qquad
\label{eq:eqO-9-8}  
 \left. \frac{dV}{dr} \right|_{r=0}=0 \;.
\end{equation}
For every choice of $Z$, $\gamma$ is determined by the boundary condition on the electrostatic potential at the origin. Eqs. \ref{eq:eqO-9-8} are therefore eigenvalue equations for $\gamma$, the unscreened part of the nuclear charge, and $Z-\gamma$ gives the charge of the vacuum:
\begin{equation}
\int d^3x\langle \Omega|\rho_e|\Omega|\rangle =e(Z-\gamma)
\;. \label{eq:eqO-9-9}
\end{equation}
Neglecting at first the inhomogeneity of the solution, we find that $V(0) = V_0$ is determined from the condition
\begin{equation}
\rho_T=\rho_N+\langle \Omega|\rho_e|\Omega|\rangle =0
\;, \label{eq:eqO-9-10}
\end{equation}
in the limit of large Z, \ie\ when the distribution of nuclear charge is large compared with l/m, then
\begin{equation}
eV_0=m-[m^2+(3\pi^2\rho_N)^{2/3}]^{1/2}\to -(3\pi^2\rho_N)^{1/3}
\;. \label{eq:eqO-9-11}
\end{equation}
Integration of \req{eq:eqO-9-7} is straightforward. An equal number of protons and neutrons and normal nuclear density have been assumed for the nuclear charge distribution. The results~\cite{[70]} for $\gamma$ are plotted in \rf{Fig_m10}. 
\begin{figure}[t]
\centering{
\resizebox{0.6\textwidth}{!}{%
\includegraphics{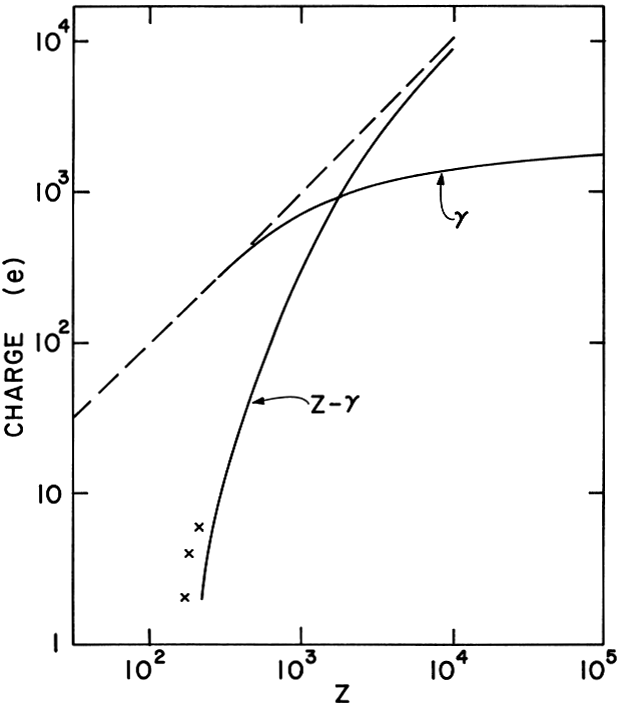}
}
}
\caption[]{The unscreened charge $\gamma$ and the total charge of
the vacuum $(Z-\gamma)$ as a function of Z. The crosses
denote points from single-particle calculations. The
dashed line denotes the nuclear charge Z.}\label{Fig_m10}
\end{figure}
From the figure, one can see that $\gamma$ increases monotonically with Z, and that $\gamma/Z$ decreases as Z increases. In fact, from the requirement that $V_0$ remains constant with growing Z, at the surface of the nuclear charge distribution we find
\begin{equation}
V_0\sim-\dsfrac{\gamma(Z)}{R(Z)}
\;, \label{eq:eqO-9-12}
\end{equation}
and, since $R(Z)\sim Z^{1/3}$
\begin{equation}
\dsfrac{\gamma(Z_1)}{\gamma(Z_2)}=\left(\dsfrac{Z_1}{Z_2}\right)^{1/3}
\;. \label{eq:eqO-9-13}
\end{equation}
The single-particle results are denoted by crosses in \rf{Fig_m10} and agree reasonably well with an extrapolation of the Thomas-Fermi results into the realm of small values of $Z-\gamma \simeq1$.  The radial total charge density, calculated from the right-hand side of \req{eq:eqO-9-7}, is shown in \rf{Fig_m11}. The results are scaled with $\gamma$ such that each curve is normalized to unity. We see that the charge density resembles more and more that of a surface dipole with clearly defined regions of positive and negative charge. This holds true since the characteristic wavelength of the electron charge is defined by $1/m_e$ while the externally prescribed \lq\lq nuclear\rq\rq\ charge distribution has a sharp edge that the electron wave functions cannot follow. For a more recent discussion of this phenomenon see Ref.\,\cite{Madsen:2008vq}.

\begin{figure}[t]
\centering{
\resizebox{0.7\textwidth}{!}{%
\includegraphics{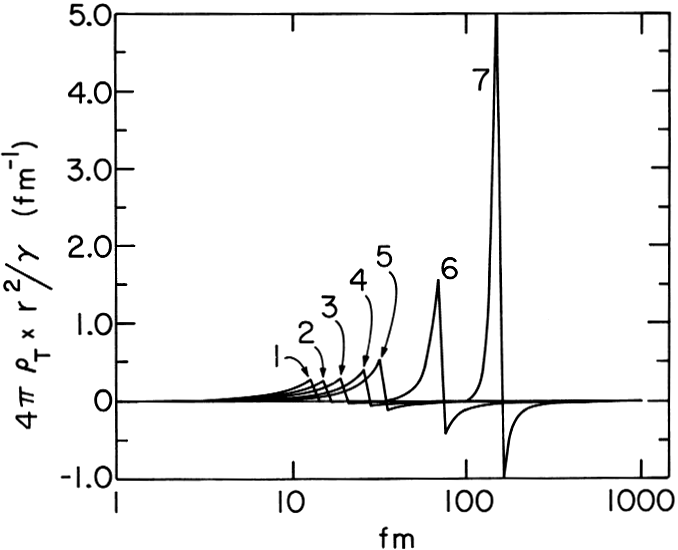}
}
}
\caption[]{The total charge densities, scaled with $\gamma$. Curves $1\equiv Z = 600;\
 2\equiv1000;\  3\equiv2000;\ 4\equiv5000;\ 5\equiv
10.000; 6\equiv 10^5; 7\equiv 10^6$.}\label{Fig_m11}
\end{figure}
The charge generated by successive levels joining the lower continuum is sufficient to screen most of the bare nuclear charge. Our results have shown that there is a limit to the coupling strength between electrons and charged matter. The boundary conditions chosen here, that of uniform density background charge, have led to the finite self-consistent potential step, $V_0$. The relevance of this discovery as originally described by M\"uller and Rafelski~\cite{[70]} is best documented by the fact that during the subsequent years it has been rediscovered several times~\cite{[119a],[119b],[120a],[120b]}, and that a detailed proof of the concept was presented in Ref.\,\cite{Madsen:2008vq}. 

\subsubsection{Self energy effects}\label{Sec334}
To close we make a few remarks about the electromagnetic self-energy corrections in high $Z$ systems. One usually writes the result in the form
\begin{equation}
\Delta E_{se}=\dsfrac{\alpha}{\pi}\dsfrac{(Z\alpha)^4}{n^3}F(Z\alpha)m \ , \ (l=0)
\;,
\label{eq:eqO-6-26}
\end{equation}
where $n$ is the main quantum number of the atomic state in question. $F(Z\alpha)$ is a function that can be obtained either as a series in $Z\alpha$ through a perturbation expansion or exactly through numerical computations employing the exact electron propagator in the external field \req{eq:eqO-4-27}. The perturbative approach was carried through by Erickson and Yennie~\cite{[78a],[78b]}. For $Z\alpha$ approaching unity, this method becomes less and less reliable and numerical calculations are called for~\cite{[79],[79A]}. Early results by Erickson~\cite{[80]} and by Desiderio and Johnson~\cite{[81]} are valid up to $Z\sim 100$. Exact results for point nuclei up to $Z = 137$ were given by Mohr~\cite{[82]}. A calculation extending beyond $Z\alpha=1$ is due to Cheng and Johnson~\cite{[83]} who made use of the eigenfunction expansion of the electron Green\rq s function for an extended nucleus. At Z = 160 they find $F(Z\alpha) = 3.34 \pm 0.16$ and a value of $\Delta E_{se} = 7.37 \pm 0.35$ keV for the energy shift of the $1s$-state, increasing with $Z$. 

It is clear that the self-energy correction reduces the total binding energy and delays the diving process. Unfortunately, Cheng and Johnson~\cite{[83]} were not able to obtain a reliable estimate of $\Delta E_{se}$ for $Z=Z_\mathrm{cr}$ due to numerical problems. A calculation of the two-loop \ie\ ${\cal O}(\alpha^2)$ irreducible contribution of the second-order electron self-energy for hydrogenlike ions with nuclear charge numbers $3\le Z\le 92$ was also presented~\cite{Goidenko:1999pa}. The interaction with the nuclear Coulomb potential is treated nonperturbatively in the coupling constant $Z\alpha$. From the perspective of our interest in the diving process these results do not introduce any new elements: it should be stressed that there is no reason to expect that the importance of the self-energy corrections should increase at the diving point. On the contrary, one may suspect that the quantum self energy approaches more and more the classical self-energy of the charge distribution of the $1s$-state (which is approximately 10 keV at $Z = 170$) as this state becomes more and more isolated from all other states. 

For a recent review of the different relativistic and QED effects at high $Z$ obtained with the help of Dirac-Fock method we refer to Ref.\,\cite{Indel07}.

\section{Heavy-ion collisions and positron production}\label{Sec4}
\subsection{Quasi molecules}\label{Sec41}
Early on it was recognized that in heavy-ion collisions the relativistic deeply bound electrons were moving fast enough to form quasi-molecular states around the two slowly moving nuclear Coulomb potential centers. This insight engendered the proposal that the collision of two extremely heavy nuclei, \eg\ U and U, could be used to probe the charged vacuum~\cite{[9],Rafelski:1972mf,[44]}. The relatively slowly moving heavy-ions with energies at the Coulomb barrier provide a common field for a shared quasi-molecular electron cloud. These electron eigenstates could be computed in a good approximation using the combined Coulomb field corresponding to a super-heavy nucleus of charge $2Z$, with a quasi potential formed by a charge distribution with diameter $2R_N=R_{12}$ corresponding to the distance $R_{12}$ between the two nuclei~\cite{Rafelski:1972mf,[23]}. 

This \lq monopole\rq\ approximation can be justified by averaging the two lowest terms in the multipole expansion \req{eq:eqO-2-6}. Adopting such an effective radial form of the potential to simulate the effect of axially symmetric potential implements the idea of quasi-molecular states where the electrons circle around the two centers, or seen in reverse,  the two nuclear charges circle around each other, and the electron is observing the so obtained averaged potential. The shape of the adopted effective monopole radial potential is seen in \req{eq:eqO-2-11}, where the nuclear radius $R_N \to R_{12}/2$. 

In \rf{Fig_m12} we compare the true and approximate forms of the potential where they differ most. The exact two center potential following the axis connecting the two nuclei (dashed line) is compared to the monopole approximate potential (solid line) for the case of a Uranium-Uranium collision. This shows that the electrons experience attractive forces similar to those of a super-heavy nucleus with $Z_\mathrm{eff}=184$ protons. This simple approximation was tested extensively using the numerical methods that were developed in Ref.\,\cite{[39]}, and found to be a very useful tool in understanding the physics of strong fields in heavy-ion collisions at sub- and near-Coulomb barrier collisions.

\begin{figure}[t]
\centering
\resizebox{0.7\textwidth}{!}{%
\includegraphics{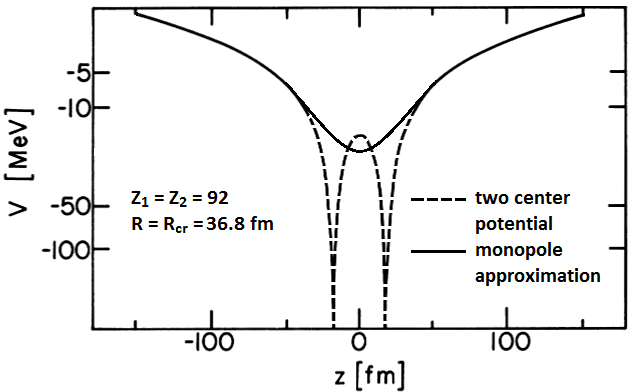}
}
\caption[]{
Solid line: the (averaged) monopole potential that can be used to compute the electron binding in presence of colliding heavy-ions, see text; dashed line: the two center potential cut along the axis connecting the two nuclei.}\label{Fig_m12}
\end{figure}

\subsection{Towards experimental observables}\label{Sec42}
The following experimentally observable effect emerges as a consequence of the supercritical binding: in collisions of high $Z$ heavy ions an empty $1s$-state can be bound by more than $2m_ec^2$. Subsequently, a positron is emitted spontaneously. When the heavy ions separate again, the previously empty $1s$-state is now occupied by an electron, thus we effectively produced a pair by spontaneous vacuum decay. The actual physical situation is not that simple: the heavy-ion collision is a time-dependent process, thus there may not always be enough time to emit a positron.

The range of collision parameters of interest is shown in the left part of \rf{Fig_m13}, where the plane $Z$--$E_\mathrm{kin(lab)}$ is depicted. The kinetic energy relates directly to the achieved distance of closest approach. It shows a lower boundary below which no spontaneous vacuum decay occurs. Moreover we note a flat domain labeled \lq Coulomb Barrier\rq\ where the nuclei will come in contact. Collisions near to this condition may favor formation of a surface-sticking nuclear quasi molecule. It has been proposed~\cite{[30]} that to prolong the time that heavy-ions spend close to each other one should explore this effect in specific nuclear collision systems. We show an illustration of this situation in \rf{Fig_m13} on the right. It is hoped that due to nuclear interactions and under certain kinematic conditions the colliding nuclei could stick to each other long enough to permit the observation of a well defined \lq peak\rq\ in the positron spectrum that is characteristic of the neutral vacuum decay in supercritical fields.

\begin{figure}[t]
\centering{
\resizebox{\textwidth}{!}{%
\includegraphics{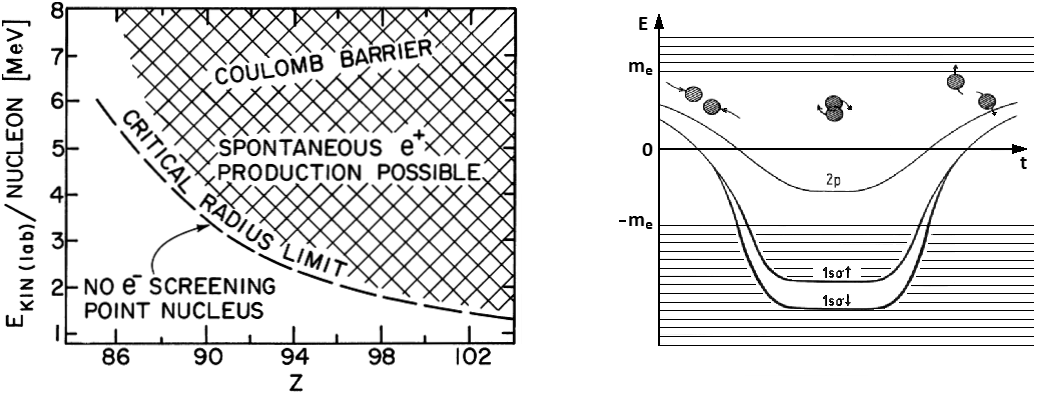}
}
}
\caption[]{On left: the plane $Z$--$E_\mathrm{kin(lab)}$ is depicted lower boundary is the critical distance, upper collision energy is bounded by nuclei running into each other (Coulomb barrier). On right: For collisions where nuclei touch each other at the Coulomb barrier it is possible that there is effective sticking time during which the spontaneous positron emission is amplified, Ref.\,\cite{[30]}. }\label{Fig_m13}
\end{figure}
 
The rather short lifetime of a supercritical K-shell vacancy against positron emission, $\tau_{e^+}\simeq 1O^{-18}$--$10^{-19}$\,sec implis that the supercritical system needs to live only for such a short period of time. It has therefore been proposed that the collision of two extremely heavy nuclei, \eg\ U and U, could be used to probe the charged vacuum~\cite{[9],Rafelski:1972mf,[44]}. An estimate of the order of magnitude shows that this is indeed feasible: the non-sticking typical collision time of two nuclei at energies just below the Coulomb barrier is
\begin{equation}
\tau_{\rm coll}\simeq \displaystyle \frac{2R_\mathrm{cr}}{v}\simeq 0.25 \times 10^{-20}\,\mbox{sec}\label{eq:eqO-3-1}
\end{equation}
with $R_\mathrm{cr}\simeq 35$\,fm (see below). The emission time for positrons is typically 100 times longer such that one expects a yield of roughly 1\% in this reaction. The theoretical treatment of the process is greatly facilitated by the large mass of the two nuclei: the Sommerfeld parameter $\eta=Z_1Z_2\alpha/v> 500$. Hence the classical approximation to the nuclear motion is adequate, and only the electrons have to be treated quantum mechanically.

Because of the similarity to stable or metastable molecules formed by valence-shell electronic binding, the binary systems described above are called quasi-molecules. The formation and existence of such inner shell quasi-molecules proposed theoretically in Ref.~\cite{[9]} has been ascertained by the observation of X-radiation from the transition between molecular states~\cite{Kirsch:1978yor,Anholt85}. Due to the collision dynamics the quasi-molecular orbitals (MO\rq s) are strongly varying in time. This can lead to electron excitations and, correspondingly, hole creation and subsequent MO X-ray emission. Theoretical predictions of these experimentally observable quantities are obtained by time dependent perturbation theory where one expands the electronic scattering states in terms of the quasi-stationary solutions of the two center Dirac equation~\cite{[39],[57a],[57b],[58]}. Of course, other basis systems are possible (\eg\ the atomic basis of the target nucleus or the projectile). However, according to the adiabaticity of electronic motion of the inner shells the most suitable approach is given by the adiabatic two center states. 

\begin{figure}[t]
\centering{
\resizebox{0.7\textwidth}{!}{%
\includegraphics{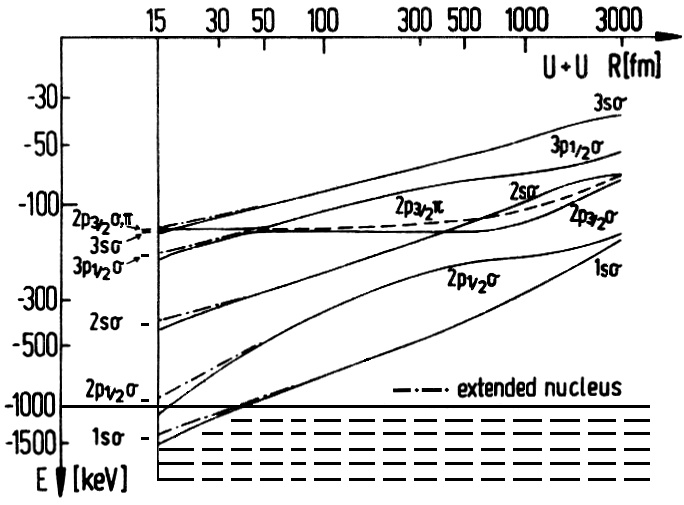}
}
}
\caption[]{The stationary quasi-molecular eigenvalues in U+U  collision as a function of two center distance $R$. Dot dashed: including nuclear size.}\label{Fig_m14}
\end{figure}

The numerical integration of \req{eq:eqO-2-7a} and \req{eq:eqO-2-7b} determines the energy eigenvalues~\cite{[59],[59a]} represented in \rf{Fig_m14} for the lowest levels in the symmetric system $^{92}$U+$^{92}$U. Comparing  with \rf{Fig_m02} we see that the  charge $Z$ is replaced by two center distance $R$ that can be changed as function of time. The influence of the nuclear extension on the molecular levels is demonstrated by the difference between the full lines (for point like nuclei) and the dashed dotted lines (for extended nuclei).

The quasi-molecular states are usually classified by the quantum number $\mu$ of the angular momentum component in the direction of the internuclear axis. $\mu$ has eigenvalues $\mu= \pm 1/2, \pm 3/2,\pm 5/2,\ldots$ which are symbolically denoted by $\sigma, \pi, \delta,\ldots$. One often assigns the quantum number of the united atomic state ($R\to 0$) to the two center wave function to which it is correlated (1s$_{1/2}\sigma$, 2s$_{1/2}\sigma$, 2p$_{1/2}\sigma$, 2p$_{3/2}\sigma, \ldots$). Since for symmetric systems the parity is also a constant of motion we can furthermore distinguish between even and odd states in this case.

The eigenstate energy of of most tightly bound electrons increases as ions approach and at $R_\mathrm{cr}\simeq 35$\,fm, it equals $-2m_e$ a for the $1s_{1/2}\sigma$ electron state. The quasi-molecule is rendered supercritical in just the same way as the super-heavy atom was at $Z> Z_\mathrm{cr}$. In further approach the finite extension of the two nuclei becomes important. However, the precise value of the critical distance is influenced by less than 1.5 fm~\cite{[60]}. Screening due to the presence of other electrons is thus equally or more important~\cite{[61]}.

For the understanding of the ionization processes the ion separation near to where levels approach each other closely are most relevant. If two levels belong to different quantum numbers they are allowed to cross. However, in the symmetric U+U case states with equal parity these states repel due to the Wigner -- von Neumann rule. The asymptotic designation ($1s_{1/2}$, $2p_{3/2}$, etc.) of $\sigma$-states becomes meaningless after such pseudo-crossing. In fast heavy-ion collisions the use of the so-called diabatic basis~\cite{[62]} where no dynamical coupling exists and all states may cross helps in evaluation of the probability of inner shell ionization, which is prerequisite to emission of positrons.

\begin{figure}[t]
\centering{
\resizebox{0.7\textwidth}{!}{%
\includegraphics{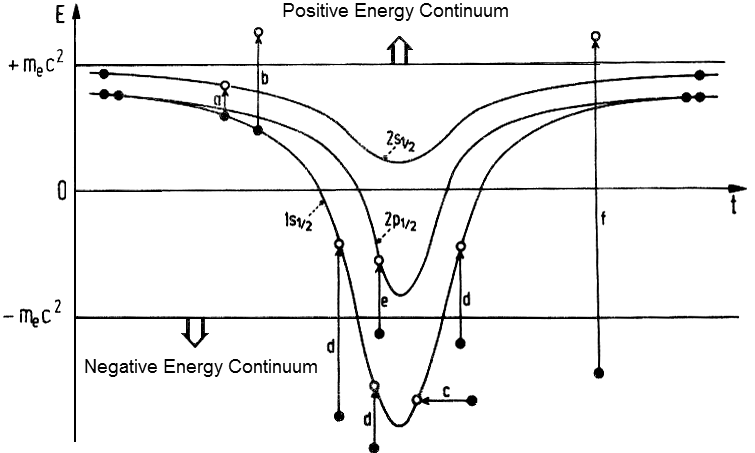}
}
}
\caption[]{Schematic representation of pair-production processes
in heavy-ion collision as a function of time. We see most tightly bound eigenstates and relevant processes: $a,b$-ionization; $c$-spontaneous
and $d,e$-induced vacuum decay, $f$-continuum pair production.}\label{Fig_m15}
\end{figure}

In order to evaluate the positron production cross section due to strong fields effects it is necessary to consider the dynamical processes that are present in a collisions event. They are depicted in \rf{Fig_m15} as a function of time: initially the binding increases but beyond the point of closest approach of the ions it decreases. For the positron production to involve the tightly bound eigenstate we need to remove electrons still present in the K-shell quasi-molecular states, see processes $a,b$. The motion of the ions can induce positron production in the processes $d,e$, there can be furthermore direct free pair production process $f$. Coherently superposed to processes $d,e,f$ is the spontaneous positron emission process $c$. Detailed calculations in the decade 1970-1981 of the theoretically anticipated effects can be found in ~\cite{[24A],[24B],[25],[26],JoReinStick81}.

We show representative examples for positron production~\cite{[26],JoReinStick81} in \rf{Fig_m16}. On left for four different systems with total charge $Z_1+Z_2=164, 174, 184, 190$ at bombarding energy 5.9 MeV/u, only the last two systems are supercritical -- the positron yield increases significantly with $Z_1+Z_2$ but there is no peaked structure as the time induced $d,e,f$-processes dominate. On right in \rf{Fig_m16} for the $Z_1+Z_2=184$ the U+U system positron production allowing for nuclear sticking time~\cite{[30]} is shown. As sticking delay $T$ grows, the decay time of the supercritical resonance begins to generate a much more intense positron line -- the lack of time dependence favors the spontaneous over induced process and the spontaneous process is more sharply peaked in energy. 

\begin{figure}[t]
\resizebox{\textwidth}{!}{%
\includegraphics[width=0.5\columnwidth]{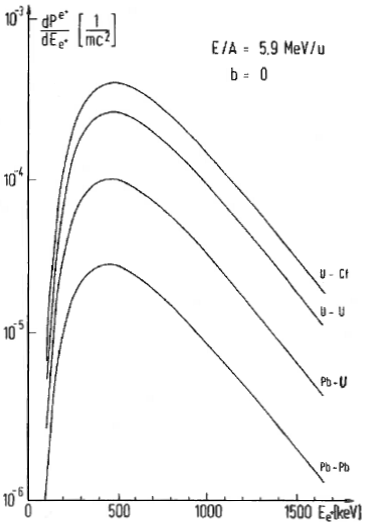}
\includegraphics[width=0.5\columnwidth]{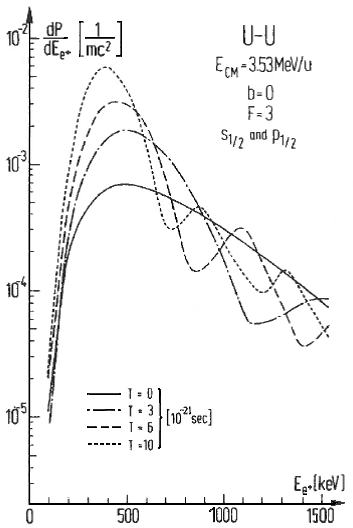}
}
\caption[]{Positron production spectra in heavy-ion collisions. On left: Coupled channel calculations for 5.9 MeV/$u$ collisions of various systems. On right: example of enhancement generated by nuclear sticking in U-U system with delay times $T\,=\,0;\,3*10^{-21};\,6*10^{-21};\,10^{-20}$s. For large sticking times $T$ a line due to spontaneous positron production emerges.}\label{Fig_m16}
\end{figure}

\subsection{Positrons from heavy-ion collisions}\label{Sec43}

Before closing let us briefly describe the experimental status of the search for spontaneous vacuum decay. A series of experiments searching for spontaneously emitted positrons was carried out over a period spanning about two decades (1977-1999). The initial experimental results on pair production in heavy-ion collisions~\cite{[26],JoReinStick81} were compatible with the predictions of strong field QED as is seen for example in \rf{Fig_m16}. In particular, the data confirmed the highly nonperturbatve nature of the positron production process with a $Z^n (n>20)$ dependence on the nuclear charge. These results, however, did not establish conclusively the novel mechanism of spontaneous pair production given the nature of the theoretically predicted positron spectra, which are dominated by the induced vacuum decay process. Thus, if the topic of strong fields QED ended with these initial experiments, we would have today indirect, but not convincing, evidence to celebrate the discovery of the structured charged vacuum in strong fields.

However, the experimental groups, being under intense internal competition, picked up the idea of nuclear sticking~\cite{[30]}. The early 1980s saw the beginning of the experimental search for sticky nuclear collision conditions possibly leading to narrow positron lines. The experimental results were at first very exciting, showing the emergence of peaked positron spectra just as predicted under optimal circumstances~\cite{Schweppe:1983yv,Kankel83,Kienle84}. 
 
Questions about this interpretation arose when the experimental groups found positrons produced in subcritical systems~\cite{Kienle87}, and soon after also electrons were seen accompanying the positrons~\cite{Cowan:1985cn,Cowan:1986fj,Salabura:1990fe}. Even so, a diligent effort was made searching for nuclear systems where possibly true vacuum decay positrons could be found~\cite{Kienle92,Konig:1992yr,Leinberger:1996pd}. All efforts were ended when improved experiments failed to find peaked positron lines where earlier experiments had seen them~\cite{APEX97,APEX99}. The consensus view today is that the earlier intriguing observations were due to highly system dependent nuclear excitations converting into pairs~\cite{Heinz:2000zw}.

The effort to interpret the data in terms of light particles decaying into pairs (MeV-mass scale axions), should be also mentioned but will not discussed further in these pages,  see~Ref.\,\cite{Widmann:1991wn,Muller:1991sg} for a pertinent discussion. The positron line interpretation based on the existence of a new elementary particle has been ruled out in the study of $e^+e^-$ resonant scattering~\cite{Widmann:1991wn}. In this context, still further effort was made to introduce composite states allowing for a form factor of the new particle~\cite{Schramm:1988np}. In such a case its production may be possible in extended domains of strong fields, but not in $e^+e^-$ reactions. It is amusing to note that the search for light neutral bosons decaying into $e^+e^-$ pairs has recently received renewed interest in the context of the search for a so-called ``dark photon''~\cite{Boyce:2012ym}.

\section{Summary}\label{Sec16}
In this review we have presented the understanding of the relativistic quantum theory and its 2nd quantization in presence of arbitrary strong external fields looking back at the theoretical work carried out before 1982. We could do this since {\em little or even nothing has changed in the theoretical formulations since}.  

In order to recall the common characteristics of all phenomena we addressed, let us summarize the basic results: when the field of force acting on any species of particles (electrons, pions, gluons, etc.) exceeds a particle related critical strength, the vacuum state is forced to change. For instance, in the original case of a supercritical atomic nucleus the vacuum state becomes charged, and positrons are emitted at the same time. In general, the vacuum state is rearranged in such a way as to diminish the effect of the applied \lq external\rq\ force, \ie\ the vacuum acts as a screening medium.

The study of QED of strong fields was one of the key developments that has facilitated the development of new ideas about the vacuum state. In our view these developments were essential for the precognition and the understanding of the true nature of the vacuum of quantum chromodynamics which followed. 

Today there is a new interest to return to the physics of QED of strong fields in a new experimental realm. In the focal point of very short pulse ultra intense lasers~\cite{Hegelich:2014tda} we approach if not today, then in the near tomorrow the critical fields condition. The ultra intense lasers just like the heavy-ion collisions draw their importance from the fact that they form the unique laboratory based testing ground for QED of strong fields. The entirely non-perturbative nature of pair-production accompanying change in the ground state allows to explore in laboratory processes in which change in the vacuum structure are turned on and off. 


\end{document}